# Point Defects and Localized Excitons in 2D WSe$_2$

*Yu Jie Zheng[1,2,†], Yifeng Chen[2,†], Yu Li Huang[1,3], Pranjal Kumar Gogoi[1], Ming-Yang Li[4], Lain-Jong Li[4], Paolo E. Trevisanutto[2], Qixing Wang[1], Stephen J. Pennycook[5,*], Andrew T. S. Wee[1,2,*], Su Ying Quek[1,2,*]*

[1]Department of Physics, National University of Singapore, 2 Science Drive 3, 117542, Singapore

[2]Centre for Advanced 2D Materials, National University of Singapore, Block S14, Level 6, 6 Science Drive 2, 117546, Singapore

[3]Institute of Materials Research & Engineering (IMRE), A*STAR (Agency for Science, Technology and Research), 2 Fusionopolis Way, Innovis, 138634, Singapore

[4]Physical Sciences and Engineering, King Abdullah University of Science and Technology, Thuwal, 23955-6900, Saudi Arabia

[5]Department of Materials Science & Engineering, National University of Singapore, 9 Engineering Drive 1, Singapore 117575

**ABSTRACT**

Identifying the point defects in 2D materials is important for many applications. Recent studies have proposed that W vacancies are the predominant point defect in 2D $WSe_2$, in contrast to theoretical studies, which predict that chalcogen vacancies are the most likely intrinsic point defects in transition metal dichalcogenide semiconductors. We show using first principles calculations, scanning tunneling microscopy (STM) and scanning transmission electron microscopy experiments, that W vacancies are not present in our CVD-grown 2D $WSe_2$. We predict that O-passivated Se vacancies ($O_{Se}$) and O interstitials ($O_{ins}$) are present in 2D $WSe_2$, because of facile $O_2$ dissociation at Se vacancies, or due to the presence of $WO_3$ precursors in CVD growth. These defects give STM images in good agreement with experiment. The optical properties of point defects in 2D $WSe_2$ are important because single photon emission (SPE) from 2D $WSe_2$ has been observed experimentally. While strain gradients funnel the exciton in real space, point defects are necessary for the localization of the exciton at length scales that enable photons to be emitted one at a time. Using state-of-the-art GW-Bethe-Salpeter-equation calculations, we predict that only $O_{ins}$ defects give localized excitons within the energy range of SPE in previous experiments, making them a likely source of previously observed SPE. No other point defects ($O_{Se}$, Se vacancies, W vacancies and $Se_W$ antisites) give localized excitons in the same energy range. Our predictions suggest ways to realize SPE in related 2D materials and point experimentalists toward other energy ranges for SPE in 2D $WSe_2$.

Elucidating the nature and properties of point defects in materials is of fundamental importance. For 2D materials, in particular, complete elimination of defects in growth processes is still a challenging task, and the high surface-to-volume ratio also renders 2D materials more susceptible to unintentional reactions with chemicals in the environment. The implications of point defects on the properties of 2D materials are multifold. It has been found that removing sulphur vacancies ($S_{vac}$) in $MoS_2$ by reaction with thiol molecules can result in much improved optical quality.[1] This observation can be attributed to the gap states in $S_{vac}$ and the fact that $S_{vac}$ are susceptible to adsorption of contaminants from the environment. On the other hand, defects in transition metal dichalcogenides (TMDs) can have a large impact on spin transport properties in graphene/TMD heterostructures.[2] Point defects, together with strain gradients, are also believed to play an important role in single photon emission (SPE) from 2D $WSe_2$,[3,4] which has been observed by several experimental groups.[5-10]

It is interesting to ask why most successful SPE experiments in TMD have utilized $WSe_2$, rather than the more common TMD, $MoS_2$. It has been suggested that the reason for this phenomenon is that $WSe_2$ hosts low-lying dark exciton states, which can increase the lifetime of the excitons and assist in the funneling of excitons to point defects. However, several theoretical predictions suggest that 2D $MoS_2$ may also host dark exciton states that are lower in energy[11] or degenerate[12] with the bright exciton. On the other hand, 2D $WSe_2$ has been found to have excellent optical quality compared to $MoS_2$.[13] These differences suggest that the defect structure in $WSe_2$ is fundamentally different from that in $MoS_2$. We remark that most of the optical experiments have been performed on exfoliated samples rather than samples grown by chemical vapour deposition (CVD), where grain boundaries can also be present. Scanning tunnelling microscopy (STM) and first principles density functional theory (DFT) calculations have shown that sulphur vacancies $S_{vac}$ are the predominant point defects in mechanically exfoliated $MoS_2$ monolayers.[14] First principles calculations have also predicted that the chalcogen vacancy ($Se_{vac}$) has the lowest formation energy among the intrinsic point defects in single layer (SL)-$WSe_2$,[15] and it is therefore natural to ask if $Se_{vac}$ could be the point defect giving rise to localized excitons in SL-$WSe_2$. However, recent electron-beam irradiation experiments on SL-$WSe_2$ have shown that the $Se_{vac}$ created by the electron beam gives a broad emission peak extending to at least 0.2 eV below the A peak,[16] while many-electron GW-BSE calculations on $Se_{vac}$ in SL-$WSe_2$[17] and $S_{vac}$ in $MoS_2$[18] predict broad optical absorption spectra arising from chalcogen vacancies. In contrast, the SPE previously observed from SL-$WSe_2$ correspond to localized excitons ~45-100 meV below the free exciton A peak.[5-10] This implies that $Se_{vac}$ are not responsible for the localized excitons in SL-$WSe_2$.[5-10] Instead, it has been suggested, based on STM images, that W vacancies ($W_{vac}$) are the predominant point defects in SL-$WSe_2$, and are responsible for localized SPE.[19]

In this work, we use a combination of first principles calculations, scanning tunneling microscopy (STM) and scanning transmission electron microscopy (STEM) experiments to investigate the point defects in SL-$WSe_2$. Our calculations show that W vacancies are not energetically favored, and the experiments show that W vacancies are not present in our CVD-grown samples. Instead, we predict that $Se_{vac}$ have the lowest formation energy among all intrinsic defects in SL-$WSe_2$. However, O-substituted Se vacancy ($O_{Se}$) and O interstitial defects ($O_{ins}$) are favored due to facile $O_2$ dissociation at $Se_{vac}$ sites at room temperature, or the use of metal oxide precursors for CVD-grown samples. Simulated STM images of these O defects agree well with our experimental STM images and the aforementioned STM results.[19] Unlike the case for $MoS_2$,[14] STM images do not reveal the presence of any chalcogen vacancies. It is interesting to note that our findings are consistent with the recent observation of O-substituted chalcogen vacancies in 2D MoSe$_2$ and 2D WS$_2$.[20] We note that the 2D MoSe$_2$ in Ref. 20 was grown from elemental sources. According to our theoretical calculations, the propensity for O$_2$ to dissociate at chalcogen vacancies is related to the work function of the 2D materials, and the energy barrier for O$_2$ dissociation at chalcogen vacancies should be largest for MoS$_2$, smallest for WSe$_2$, and intermediate for MoSe$_2$ and WS$_2$. Our results therefore contribute to a unified picture of the properties of chalcogen defects in TMD semiconductors.[21]

The passivation of Se$_{vac}$ gap states explains the superior optical quality of 2D WSe$_2$, compared to 2D MoS$_2$, and facilitates the observation of SPE from other defects. We use state-of-the-art many-body perturbation theory (GW-BSE calculations) to study the nature and energies of excitons at the experimentally observed point defects. This theoretical study is motivated by the recent SPE experiments in 2D WSe$_2$, and also by the fact that identifying the source of SPE is important, but experimentally challenging. We find that only O$_{ins}$ gives rise to localized excitons within the energy range for SPE previously observed in SL-WSe$_2$.[5-10] Our experimental investigations also reveal the presence of Se$_W$ antisite defects, which is consistent with our theoretical predictions for typical Se-rich growth conditions. Like W$_{vac}$,[17] Se$_W$ gives rise to many sub-bandgap exciton peaks that are much lower in energy than the A peak.

Taken together, we have used a combination of theory and two different experimental techniques to elucidate the nature of point defects in WSe$_2$.[21] We show that the chalcogen vacancies in SL-WSe$_2$ are passivated by atomic O, in contrast to sulphur vacancies in SL-MoS$_2$, explaining the superior optical quality of WSe$_2$, compared to mechanically exfoliated MoS$_2$ samples. We also predict the optical properties of the experimentally observed point defects, which not only shed light on recent SPE experiments, but also suggest ways to create quantum emitters in related 2D materials as well as point experimentalists toward other energy ranges for SPE in 2D WSe$_2$.

**RESULTS AND DISCUSSION**

Fig. 1a shows the formation energy of intrinsic defects in SL-WSe$_2$ on graphite, computed using density functional theory (DFT; see Methods and SI). Subscripts 'ad', 'vac' and 'int' refer to adatoms, vacancies and interstitials, respectively. 2Se$_{vac}$ refers to a double Se vacancy at the same site and Se$_W$ refers to an antisite defect with Se substituting W. As Se and W adatom and interstitial defects can be removed by annealing, they are not our focus here. Of all the intrinsic point defects we have studied for SL-WSe$_2$, we find that for both W-rich and Se-rich conditions, Se$_{vac}$ has the lowest formation energy (2.2-2.8 eV), while the defects with the next smallest formation energies are Se$_W$, W$_{vac}$ and 2Se$_{vac}$, with formation energies greater than 3.6 eV. Based on these defect formation energies, we can conclude that Se$_{vac}$ would by far be the most abundant intrinsic point defect in SL-WSe$_2$. In Fig. S1, we provide information on intrinsic point defects with formation energy greater than 6 eV. The atomic structures of all intrinsic defects are shown in Fig. S2. It is interesting to note that our predictions on the formation energy for intrinsic point defects in graphite-supported SL-WSe$_2$ are in close agreement with those for isolated SL-WSe$_2$ (Table S1). This is attributed to weak interactions between WSe$_2$ and graphite, and the fact that the strain on graphite-supported WSe$_2$ is very small (0.6%). Except for the Se adatom, gap states are evident in the densities of states (DOS) of all intrinsic defects (Fig. S3). The DOS computed using the hybrid HSE06 exchange-correlation functional is shown for Se$_{vac}$, Se$_W$ and W$_{vac}$ in Fig. 1b-d.

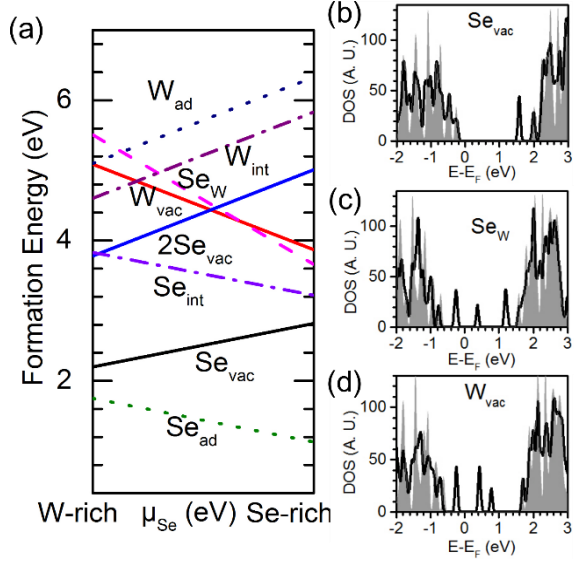

Figure 1. Intrinsic defects in SL-WSe$_2$. (a) Formation energies of the intrinsic defects for SL-WSe$_2$ on graphite. Solid lines denote vacancies, dashed lines substitutional (antisite) defects, dotted lines adatoms, and dotted-dashed lines intercalated atoms. (b-d) DOS of the 5×5 WSe$_2$ supercell with intrinsic defects. The DOS plots are computed using the HSE06 hybrid exchange-correlation functional. Gray shading: DOS of perfect WSe$_2$, aligned using the 1s core level of the W atoms furthest from the defect.

In the above discussion, we have established from our DFT calculations that just like in other TMD semiconductors,[15] chalcogen vacancies are the predominant intrinsic point defect in SL-WSe$_2$, and that all the intrinsic point defects considered here have gap states. However, as we shall show below, scanning tunnelling spectroscopy (STS) shows that all the point defects in our scanning tunnelling microscopy (STM) images have no gap states, an observation that is consistent with previous reports.[19] On the other hand, 2D materials are susceptible to unintentional chemical interactions with the environment. In ambient conditions, one of the most likely sources of contamination is molecular oxygen, O$_2$. Our DFT calculations show that the most stable of the O-related defects are dissociated O$_2$ bound at the 2Se$_{vac}$ site, O bound to Se$_{vac}$ (O$_{Se}$; Fig. 2d), O in an interstitial site within the WSe$_2$ lattice (O$_{ins}$; Fig. 2f), and O adsorbed on Se (O$_{ad}$; Fig. 2h). All of these defects have no gap states, in contrast to the intrinsic defects, as well as hydrogen and carbon extrinsic defects that we have also considered. The adsorption energy E$_{ads}$ of O on Se$_{vac}$, referenced to atomic O is ~ -7.1 eV, while that for O$_{ins}$ and O$_{ad}$ are -2.9 eV and -2.4 eV, respectively. The corresponding adsorption energies relative to molecular O$_2$ are -1.6 eV, -0.3 eV and 0.2 eV, respectively. Here, the adsorption energy relative to molecular O$_2$ tells us about the thermodynamic stability of the O-related defect relative to gas phase O$_2$ in ambient conditions, and is an indication of the thermodynamic driving force of an O$_2$ dissociation reaction. The adsorption energy relative to atomic O, on the other hand, tells us the thermodynamic stability of the O-related defect after it has formed, in the absence of other chemical species. We have also considered O bound to W$_{vac}$, Se$_W$, as well as O$_2$ bound to W$_{vac}$ and Se$_W$, and in all these defects, the formation energy is at least 2.3 eV larger (less stable). We note that our identification of the most stable O-related defects is consistent with previous literature.[22]

Next, we show that kinetically, O$_2$ can dissociate easily at Se$_{vac}$ sites in SL-WSe$_2$, resulting in O$_{Se}$ and O$_{ins}$, the most stable O-related defects. An indication of the ease of O$_2$ dissociation at chalcogen vacancies is given by the extended O-O bond lengths in O$_2$-Se$_{vac}$ and O$_2$-2Se$_{vac}$ (> 20% and > 90 % larger than its gas phase value, respectively). Using the climbing image nudged elastic band method,[23] we compute an O$_2$ dissociation barrier of 0.52 eV at Se$_{vac}$ (Fig. 2a). The reaction prefactor is computed using harmonic transition state theory,[24] giving a rate constant of ~10$^5$ counts/second at 300 K. This is consistent with the rule of thumb that a thermally activated process with barrier of ~30 k$_B$T (0.7 eV) or below can take place at room temperature for typical attempt frequencies.[25] Zero-point energy corrections reduce the barrier height by 0.04 eV. Our results indicate that $O_2$ can readily dissociate at $Se_{vac}$ at room temperature. In contrast, we compute a barrier of 0.94 eV for $O_2$ dissociation at $S_{vac}$ in SL-$MoS_2$, corresponding to a rate constant of only ~$10^{-2}$ counts/second at 300K. Fig. 2b-c shows that when $O_2$ is adsorbed at $Se_{vac}$, electrons are added to the antibonding orbital of $O_2$ and depleted from the bonding orbital and nearby W atoms, thus facilitating $O_2$ dissociation. This net charge transfer to $O_2$ (~0.6 $e^-$/atom) is in turn facilitated by the smaller work function of 2D $WSe_2$ (4.69 eV) compared to $MoS_2$ (5.06 eV). Based on the work functions, we expect the corresponding barriers for $WS_2$ and $MoSe_2$ to be larger than in $WSe_2$ but smaller than in $MoS_2$, a phenomenon that has also been observed in Ref. 26. Indeed, oxygen-substitutional defects have also been identified in $MoSe_2$ and $WS_2$,[20] while sulphur vacancies have been experimentally observed in $MoS_2$[14]. We note that quantitative prediction of energy barriers can be challenging for DFT. However, this does not affect our conclusion that $O_2$ can dissociate easily at $Se_{vac}$ sites in SL-$WSe_2$, while the same cannot be said for $S_{vac}$ in SL-$MoS_2$. Other exchange-correlation functionals also give a barrier of at most 0.57 eV (see SI, Table S3).

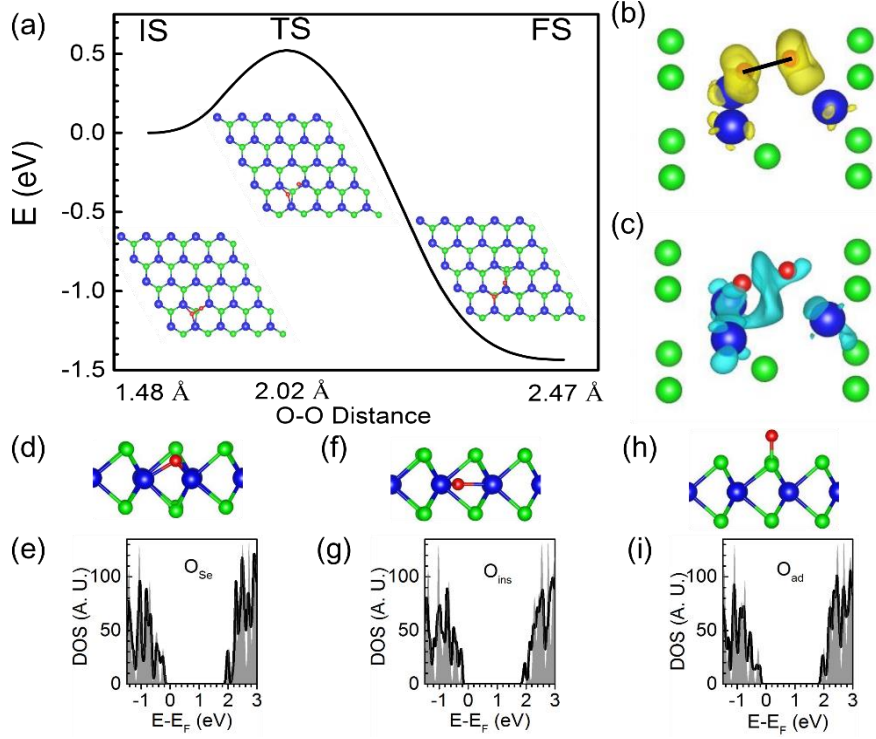

Figure 2. $O_2$ dissociative adsorption at the Se vacancy site and O-related defects in a 5 ×5 supercell. (a) $O_2$ dissociation barrier. IS, TS and FS are initial, transition and final states, respectively. (b) Electron gain and (c) loss regions $\Delta\rho$ for the IS, computed using $\Delta\rho = \rho_{IS} - \rho_{O_2} - \rho_{Se_{vac}}$. (d-i) Atomic structures and DOS (HSE06) of O-related defects. (d-e) $O_{Se}$, (f-g) $O_{ins}$ and (h-i) $O_{ad}$. Red: O, Blue: W, Green: Se. Gray shading: DOS of perfect $WSe_2$, aligned by the 1s core level of the W furthest from O.

In the final state (Fig. 2a), one O atom O1 takes the place of the missing Se (forming $O_{Se}$), while the other, O2, binds to the neighbouring W atom. The O atom O2 can move to these more stable sites $O_{ins}$ and $O_{ad}$ with thermal annealing (*e.g.* Fig. S5). None of these defects have deep gap states (Fig. 2e, 2g, 2i). Since $Se_{vac}$ is the most abundant intrinsic point defect, and $O_2$ dissociation at $Se_{vac}$ is facile, we expect the highly stable $O_{Se}$ to be the most abundant point defect in 2D $WSe_2$.

In the above discussion, we have evaluated the possibility of O defects arising from $O_2$ dissociation at chalcogen vacancies. This process is relevant to mechanically exfoliated samples, often used in optical and single photon emission experiments, as well as samples grown from elemental sources. The fact that O passivated Se vacancies were recently observed in 2D $MoSe_2$ grown from elemental sources under ultra-high vacuum conditions is consistent with our theoretical predictions.[20] On the other hand, CVD growth processes sometimes make use of metal oxide precursors, and it is relevant to consider the formation energy of the O defects under such conditions, using the chemical potential of O from the its reservoir $WO_3$. Our calculations show that $O_{Se}$ defects have the smallest formation energy (much smaller than that of $Se_{vac}$). $O_{ins}$ has a ~1 eV smaller formation energy than $Se_{vac}$ under typical Se-rich conditions (as is the case for our samples), and a comparable formation energy to $Se_{vac}$ under W-rich conditions (Table S1). Therefore, the $O_{Se}$ and $O_{ins}$ defects are likely to form *in situ* during CVD growth processes involving metal oxide precursors. We have also computed in a similar manner the formation energies of $O_S$ and $O_{ins}$ defects with those of S vacancies in $MoS_2$ monolayers, where we find that the $O_S$ defect is more stable than $S_{vac}$. However, most optical experiments are performed on exfoliated samples, and evidence of $S_{vac}$ was found in exfoliated $MoS_2$ monolayers.[1,14]

We now present an experimental characterization of the point defects in SL-$WSe_2$. Our low temperature STM measurements reveal the presence of three most commonly observed point defects ($D_1$, $D_2$ and $D_3$; Fig. 3a-c) in SL-$WSe_2$, grown *via* chemical vapour deposition (CVD) on graphite[27] and annealed in vacuum to ~300 ˚C. The STM images for $D_1$ are similar to those in Ref. 19. STM images simulated within the Tersoff-Hamann approximation[28] for $O_{Se}$, $O_{ins}$ and $O_{ad}$ compare well with the images of $D_1$, $D_2$ and $D_3$ (see Methods). Specifically, the detailed patterns of the defects are very different for different bias voltages, as revealed by the bias-dependent STM images recorded at the same area. For instance, $D_1$ exhibits as a trefoil-like pattern at $V_{tip} = 1.3$ V, but shows up as a pit-like structure at $V_{tip} = 1.2$ V, $-1.2$ V and $-1.3$ V; $D_2$ exhibits three-lobe-like features without a bright central protrusion at $V_{tip} = 1.4$ V, but a bright central protrusion is present at $V_{tip} = -1.2$ V and $-1.3$ V. No gap states are observed in the STS spectra recorded at $D_1$-$D_3$ (Fig. 3d-f). To elucidate the origin of each defect, we first need to understand their locations by superimposing pristine $WSe_2$ atomic structures to the STM images. It is worth noting that the bright spots in the STM images for the pristine $WSe_2$ can be attributed to Se sites (*e.g.*, the 1$^{st}$ row in Fig. 3b and Fig. 3c) or hollow sites (the 2$^{nd}$ row in Fig. 3a) depending on the bias voltages (Fig. S6). As a result, we can determine that $D_1$ and $D_3$ are located at the Se position, as shown in the 2$^{nd}$ row in Fig. 3a ($V_{tip} = 1.2$ V) and in Fig. 3c, respectively, while $D_2$ is located at the hollow site, as shown in the 1$^{st}$ row in Fig. 3b ($V_{tip} = 1.4$ V). The defects ($D_1$ and $D_3$) located at Se sites could be $Se_{vac}$, $O_{Se}$ or $O_{ad}$, while the one ($D_2$) located at the hollow site could be $O_{ins}$, $Se_{int}$ or $W_{int}$ (Fig. S8). By comparing the simulated STM images (Figs. 3 and S8) and the experimental STM bias-dependent images, we find that $O_{Se}$, $O_{ins}$ and $O_{ad}$ compare well with $D_1$, $D_2$ and $D_3$ respectively.

Thus, the detailed patterns observed in the STM images for $D_1$, which are very different for different bias voltages, strongly support the conclusion that $D_1$ is $O_{Se}$. The specific patterns in $D_2$, together with the presence of $O_{Se}$, further supports the conclusion that $D_2$ is $O_{ins}$. We note that $O_{ins}$ and $O_{Se}$ (at the bottom Se site) have similar STM images for low bias voltages. However, the STM images differ for larger bias voltages, and a comparison shows clearly that $D_2$ corresponds to $O_{ins}$ (Fig. S7). This is further supported by the different scanning tunnelling spectroscopy (STS) spectra for $D_1$ and $D_2$. None of the STM images simulated for the intrinsic defects (Fig. S8) match well with the experimental STM images for $D_1$ and $D_2$. Furthermore, similar to the reports in Ref. 19, no gap states are observed in the STS spectra (Fig. 3d-f). This is in contrast to the gap states predicted for intrinsic point defects and strengthens our conclusion that the defects observed here are not intrinsic point defects. Interestingly, $D_1$ ($O_{Se}$) is observed to be the most abundant point defect, as we predict for $O_{Se}$, while $D_2$ ($O_{ins}$) was also commonly observed. We were unable to locate any point defects with gap states in the STM/S experiments. This observation is consistent with the fact that the most abundant intrinsic point defect with gap states ($Se_{vac}$) would be passivated by O, removing the gap states, and resulting in O-related defects with no gap states. Other than $Se_{vac}$, the remaining intrinsic point defects with gap states have higher formation energies and are expected to occur at significantly lower densities.

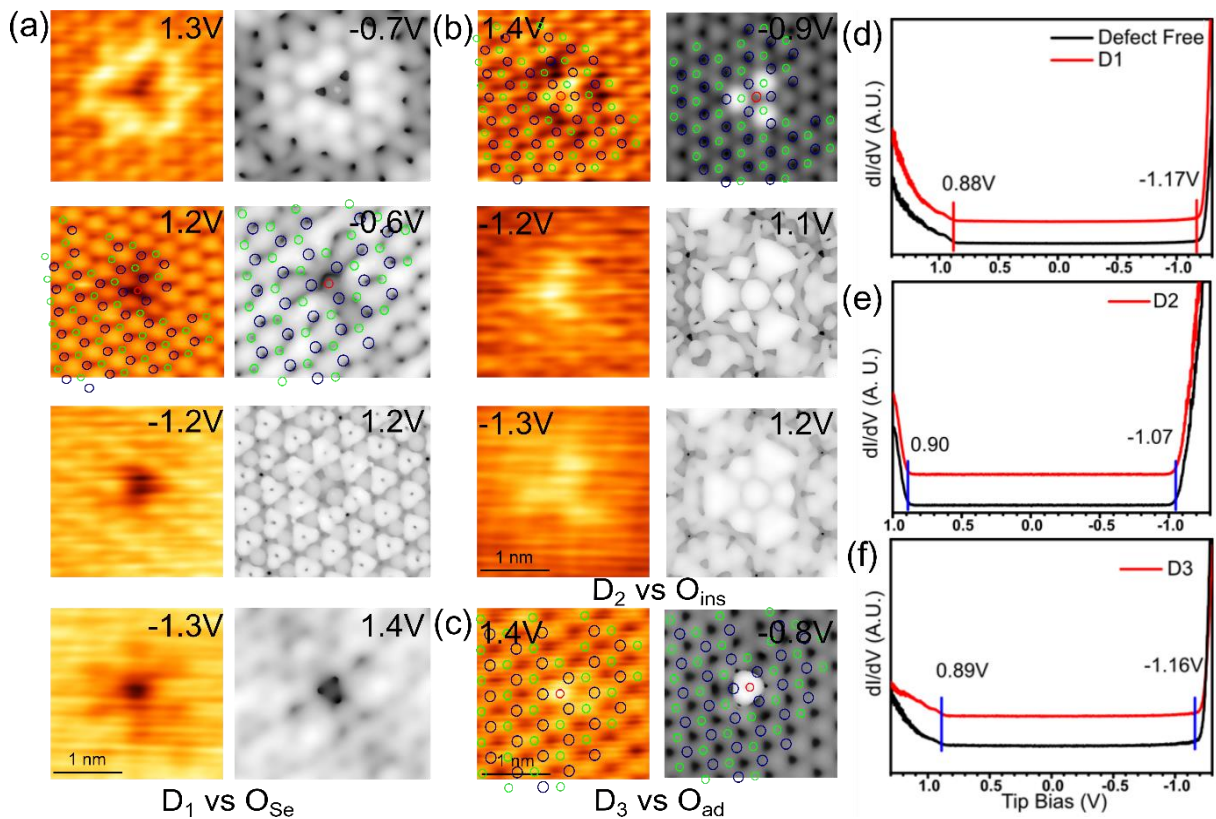

Figure 3. Experimentally observed point defects. (a-c) Left: Experimental STM images for the dominant point defects ($D_1$, $D_2$ and $D_3$) observed in STM. Right: Simulated STM images of $O_{Se}$, $O_{ins}$ and $O_{ad}$. Voltages on the left are tip bias voltages (negative biases correspond to unoccupied states and *vice versa*). Voltages on the right are relative to $E_F$ in the calculation, chosen to approximate the energy ranges of states contributing to the measured STM current. Atomic positions are shown in a, b and c. (Red: O; Blue: W; Green: Se). (d-f) Averaged STS spectra on (red) and away from (> 2nm, black) the defects.

STEM measurements were also performed on SL-$WSe_2$ (see Methods). Analysis of the STEM image intensities (Fig. S9) led to the conclusion that the most abundant defect observed was $Se_{vac}$, while $2Se_{vac}$ and $Se_W$ were also present (Fig. 4). As the atomic number of O atoms is much smaller than that of Se and W, and the HAADF STEM image intensity is approximately proportional to the square of the atomic number, O atoms are very challenging to detect above the background signal. Also, the knock-on damage threshold for the oxygen atoms may be below the 60 kV accelerating voltage used, which can also explain why O was not detected. No $W_{vac}$ were observed, consistent with typical Se-rich growth conditions.

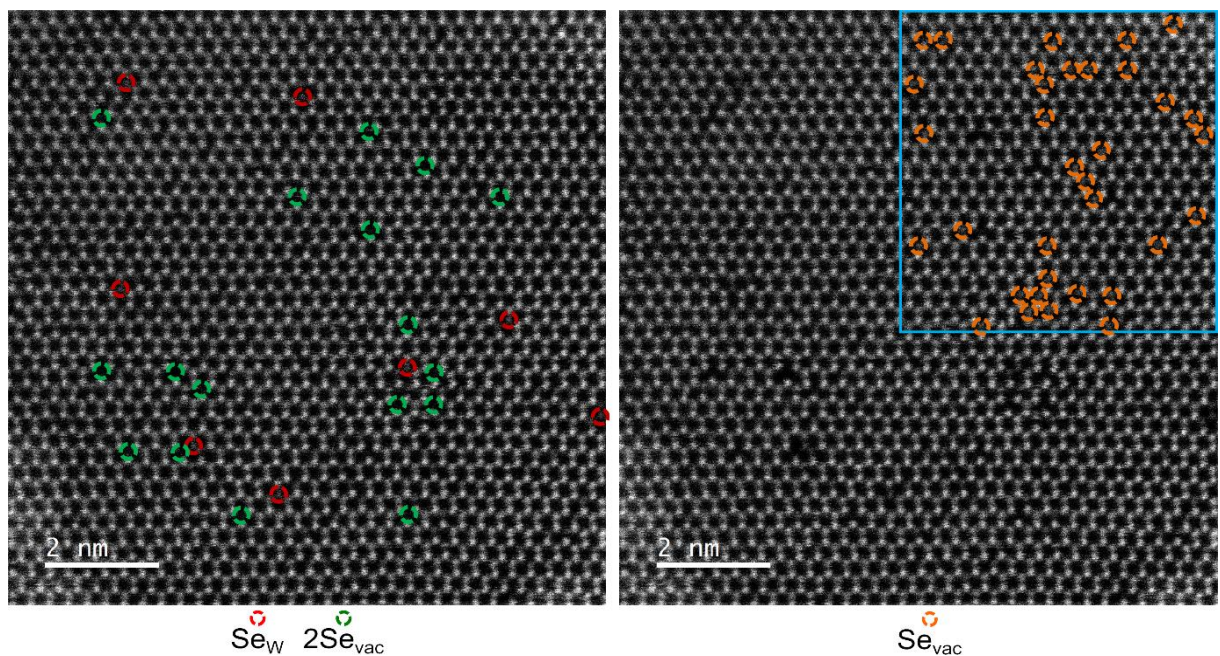

Figure 4. HAADF STEM images (60 kV accelerating voltage). The two images are for the exact same area. Se$_W$ (red) and 2Se$_{vac}$ (green) are marked in a, while Se$_{vac}$ (orange) are marked in the top right of b. O atoms cannot be distinguished due to the low knock-on damage threshold and much smaller atomic number of oxygen compared to Se and W.

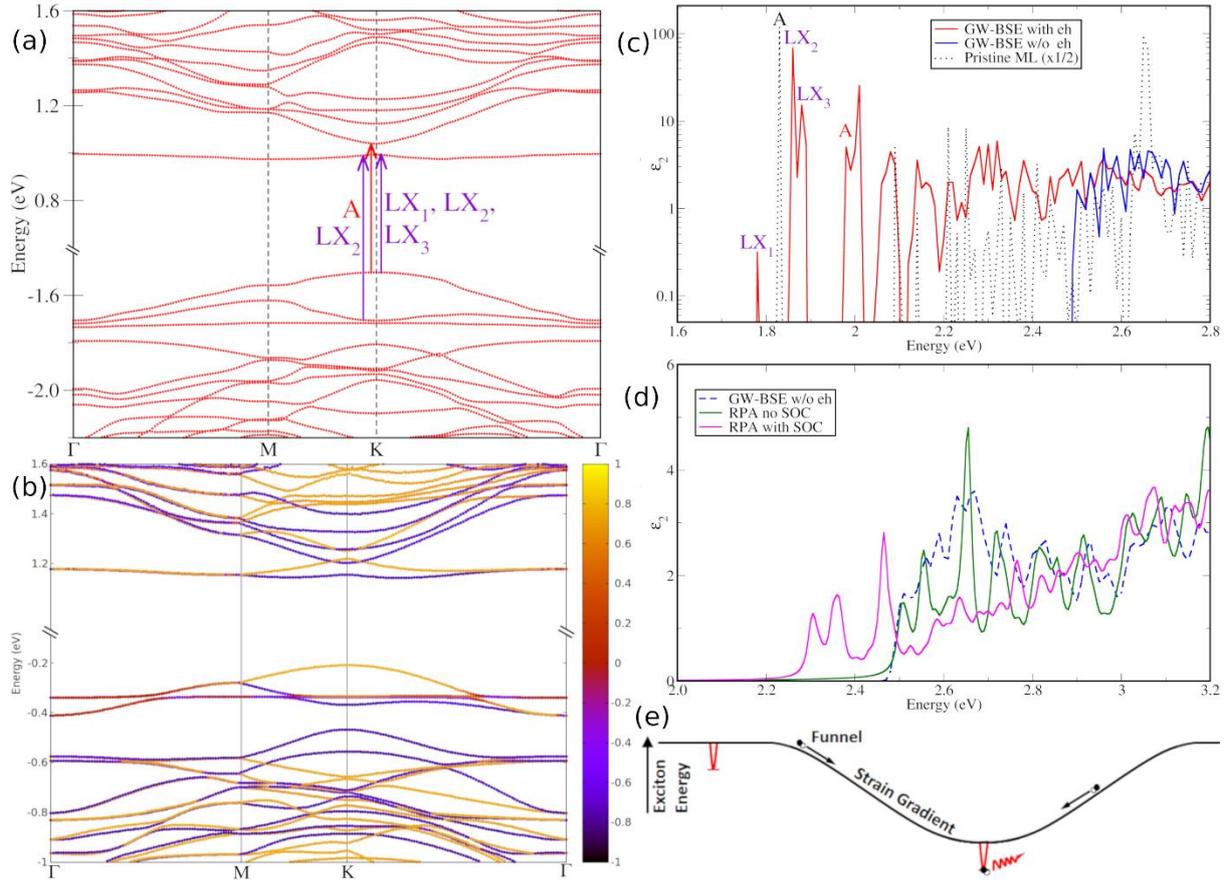

Figure 5. Electronic and optical properties for the $O_{ins}$ point defect in a 4×4 supercell of SL-WSe$_2$. (a) GW band structure, (b) DFT band structure with spin-orbit coupling (SOC) effects (the colour scale depicts the $m_z$ contributions for the states), (c) GW-BSE optical absorption spectrum, (d) RPA spectra with and without SOC effects, (e) Schematic showing that SPE arises from a combination of exciton funneling due to strain gradients, and exciton localization due to localized defect states such as in $O_{ins}$. Note that in (c) the broadening is reduced into 0.002 eV with vertical axis in log-scale to emphasize the $LX_1$ peak. The dashed line in (c) is the GW-BSE spectrum for pristine SL-WSe$_2$ with no defects. The blue line represents the optical spectrum computed in the independent particle approximation (no electron-hole interactions) using the GW eigenvalues. In (d), the RPA spectrum without SOC was shifted by +0.87 eV to align the optical onset with the GW-BSE spectrum without electron-hole interactions (due to the difference in single-particle energies between the two calculations). The RPA SOC spectrum was shifted by the same amount for comparison.

We use state-of-the-art first principles GW-BSE calculations to predict the implications of these point defects on the optical response of SL-WSe$_2$. We focus here on O$_{Se}$, O$_{ins}$, and Se$_W$. The optical properties of Se$_{vac}$ and W$_{vac}$ have been studied in the literature.[17] The GW-BSE calculations account for electron self-energy effects in the quasiparticle spectra as well as electron-hole interactions in the excitons. State-of-the-art non-uniform k-point interpolation methods[18,29] are used to obtain good convergence with respect to k-point sampling. 4×4 supercells are considered for these calculations, while optical properties are also computed in the random phase approximation (RPA) for 7×7 supercells as a comparison. Fully relativistic RPA calculations as well as GW-BSE calculations with spin-orbit coupling (SOC) added perturbatively, are also performed for 4×4 cells to estimate the effects of SOC on the optical spectra. We remark that our focus in this manuscript is on the low-energy excitons, with energies comparable to or smaller than that of the free exciton A peak. For these excitons, we expect our optical spectra to be converged.

We first focus on the O$_{ins}$ defect (Fig. 5). Interestingly, there is a low energy exciton (LX$_1$; Fig. 5c) at 1.78 eV, 50 meV below the energy of the free exciton A peak in pristine WSe$_2$. Analysis of the exciton wavefunctions (Table S5) shows that the A peak in the defect cell is at 1.97 eV (Fig. 5c), and that LX$_2$ and LX$_3$ are also defect-related excitons 30 meV and 50 meV above the energy of the A peak in pristine WSe$_2$, respectively. The difference in energy between LX$_1$ and the A peak lies in the same range as that experimentally observed in SL-WSe$_2$ (~45-100 meV), where the A peak in pristine WSe$_2$ is relevant in the limit of low defect densities.[5-10]

We note here that not all localized excitons are going to give rise to SPE. Instead, SPE is widely believed to arise from localized defects that are located in regions with tensile strain (which reduces the quasiparticle gap), funneling the electron and hole pair to the region of the defect[4,30] (Fig. 5e). Experimentally, it has been found that ~2.1% applied strain red-shifts the exciton peak by 0.1 eV in SL-WSe$_2$.[31] The slight red shift of the exciton peak with strain is also consistent with recent calculations.[32] We expect that for the samples that have not been strained deliberately,[5-10] similar amounts of strain would be present in the region of the quantum emitters, red-shifting the localized exciton peaks by ~100 meV, compared to the pristine case. This red-shift does not change our conclusions that the LX excitons in O$_{ins}$ lie in the energy range of experimentally observed single photons.[5-10] In particular, as we discuss below, no other point defects considered here give similar spectra.

For single photon emission, it is important that the exciton is localized. Indeed, we see that LX$_1$ primarily results from a combination of the bulk VBM state and defect CBM state at the K point (Fig. 5a; Table S5; Fig. S10), while other LX excitons are related to transitions between the VBM/VBM-1 states and defect CBM state. The localized nature of the defect state supports the emission of single photons, while the involvement of the bulk state is consistent with the valley polarization[18] of single photons observed in Ref. 5. From Fig. 5c, we note that electron-hole interactions are important for reducing the optical onset and creating separate energy peaks (LX *versus* A), pointing to the importance of excitonic effects in this defect system.

We further remark that screening from the substrates should not have a significant effect on the exciton peak positions. In general, screening from the substrates reduces the quasiparticle gap and the exciton binding energy by a similar amount. Since the optical gap is the difference between the quasiparticle gap and exciton binding energy, the effect of substrates on the optical peak positions is much smaller, as shown previously in MoSe$_2$[33] and MoS$_2$.[34] Furthermore, the fact that different substrates (SiO$_2$/Si,[5-7,9] hexagonal boron nitride/graphene[8,10]) give similar SPE peak positions also shows that the effect of substrate on the SPE peak positions is negligible.

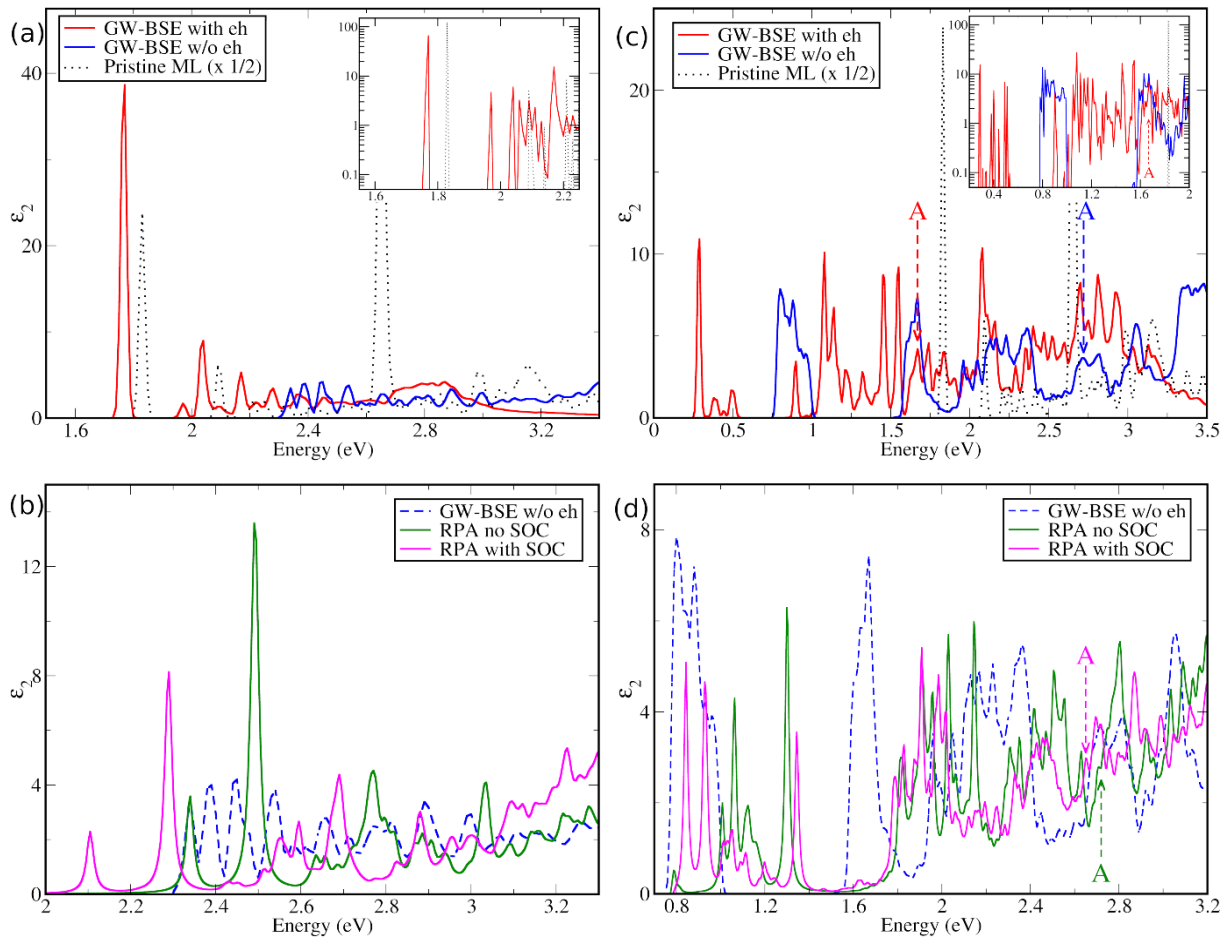

Figure 6. Optical absorption spectra for O$_{Se}$ and Se$_W$ point defects, each in a 4×4 supercell of SL-WSe$_2$. (a) GW-BSE and (b) RPA spectra for O$_{Se}$, (c) GW-BSE and (d) RPA spectra for Se$_W$. The RPA spectra in (b) and (d) are shifted by +0.92 and +0.77 eV, respectively, to match the GW-BSE spectra without electron-hole interactions. Labels A in (c) and (d) mark the free exciton A peak energy positions in the respective spectra. Insets in (a) and (c) show the low energy GW-BSE spectra plotted with a log-scale and smaller broadening.

In Fig. 6, we show the optical absorption spectra computed for $O_{Se}$ and $Se_W$ defects. For the $O_{Se}$ defect system, there are no lower energy excitons below the first prominent peak, which corresponds to the free exciton A peak and is not a localized exciton. The optical band gap is narrowed to 1.76 eV, 60 meV lower than that of pristine SL-WSe$_2$. For the $Se_W$ defect, the deep defect states (Fig. 1c) result in a large number of bound excitonic transition levels below the A peak, spanning a wide energy range (~1.56 eV), down to ~0.3 eV (Fig. 6c). This implies that $Se_W$ defects cannot be responsible for the SPE previously observed in SL-WSe$_2$. However, we predict that $Se_W$ point defects are promising for SPE in the infrared and near-infrared spectral range.

We now consider the effect of spin orbit coupling (SOC) on our conclusions. The DFT band structures of the defect supercells, including SOC effects, are shown in Fig. 5b and Fig. S11-12. We first estimate the effects of SOC by comparing DFT RPA spectra (independent particle approximation) with and without SOC (Fig. 5d, 6b, 6d). There is no obvious change in the optical activity of the transitions and the two spectra are in general very similar, except for a narrowing of the optical onset by ~0.2 eV because of the smaller DFT band gap when SOC is included. To quantify the effect of SOC on the difference in energy between LX$_1$ and the A peak in the pristine cell, we include the effect of SOC perturbatively in our GW-BSE calculations.[35] We note that the $m_z$ projection values for the $O_{ins}$ band structure near the band edges (as shown in Fig. 5b) are still large at ~0.75. These projection values are similar to that in pristine WSe$_2$ (Fig. S17a), where full spinor GW-BSE calculations and GW-BSE calculations without SOC yield comparable oscillator strengths for the A peak (Fig. S17c). A fully converged GW-BSE calculation with the perturbative SOC approach results in an A peak at 1.6 eV and a B peak at 2.0 eV in pristine SL-WSe$_2$, close to the experimentally measured 1.75 eV and 2.20 eV (Fig. S17b).[36] With the perturbative SOC procedure, the LX$_1$ peak is 16 meV above the A peak in pristine SL-WSe$_2$. Together with the effect of strain on the SPE as discussed above, we obtain an LX$_1$ peak position of ~84 meV below the pristine A peak in this approximation. The GW-BSE spectra with SOC included perturbatively are shown in Fig. S13 for O$_{ins}$, O$_{Se}$, and Se$_W$ defects. Even with the inclusion of SOC, O$_{ins}$ is much more likely than any other defect to be responsible for SPE within the energy range previously detected in experiment. The fact that O$_{ins}$ has found to exist in significant concentrations compared to other point defects with localized excitons further strengthens our conclusion.

In Fig. S14, we compare the RPA spectra for 4×4 and 7×7 supercells. For O$_{ins}$ and O$_{Se}$, the RPA spectra for 4×4 and 7×7 supercells are very similar, apart from a small shift in energy. For Se$_W$, the 7×7 supercell has fewer peaks in the RPA spectrum, but these peaks are still deep in the band gap and thus do not change our qualitative results that Se$_W$ is likely to emit in the infrared range. From Fig. S16, we also see that the DFT PBE band structures of the O$_{ins}$ system are very similar for 4×4 and 7×7 supercells.

**CONCLUSIONS**

We have presented a detailed study to identify the point defects in SL-WSe$_2$, and to predict, using state-of-the-art calculations, the nature and energies of excitons at these defect sites. Our DFT and STM results show that O$_{Se}$ is the most abundant defect in 2D WSe$_2$, passivating the gap states. This explains the superior optical quality of WSe$_2$, and facilitates the observation of SPE from localized excitons. O-substituted chalcogen vacancies have also been observed in MoSe$_2$ and WS$_2$,[20] but not in mechanically exfoliated MoS$_2$, where S vacancies with no O have been observed.[14] These findings are also consistent with our theoretical predictions and contribute to a unified picture of the interaction of oxygen with chalcogen vacancies in TMDs. We further identify $O_{ins}$ and $Se_W$ to be point defects present in 2D $WSe_2$.

The identity of point defects in 2D materials is important for many applications. In this work, we have predicted the implications of the experimentally observed point defects on the optical response. Our calculations are motivated by the SPE experiments reported in SL-$WSe_2$,[5-10] and also by the fact that it is currently challenging experimentally to ascertain unambiguously the nature of the point defect responsible for SPE, thus underscoring the importance of first principles calculations to provide key insights and guide experimentalists in future studies. The GW-BSE calculations employed here are state-of-the-art, converged with the best available k-point interpolation methods,[29] and comparable in quality with recent high quality GW-BSE calculations for defects in 2D materials.[18] We have also discussed the implications of strain, supercell size, spin-orbit coupling and substrate screening on our conclusions. The key point here is that of all the experimentally observed point defects, only one candidate ($O_{ins}$) is available that gives localized excitons in the energy range observed in recent single photon emission experiments.[5-10] Small numerical shifts that arise from strain or other effects do not change this conclusion. Our conclusion is also consistent with the fact that only $O_{ins}$ has a flat defect band close to the band edge of $WSe_2$, in contrast to other candidate point defects, which have defect states deep in the band gap, thus giving rise to photon energies that are much different from that observed in the experiments.[5-10] We emphasize that based on our results, the previous observations[5-10] of single photon emission from $WSe_2$ are related to the superior optical quality of $WSe_2$ due to the passivation of gap states by O, the presence of a flat $O_{ins}$ defect band close to the $WSe_2$ band edges, as well as the presence of strain gradients for exciton funneling. Low-lying dark excitons could also help increase the lifetime of excitons to facilitate the funneling process. These predictions suggest ways to create quantum emitters in other semiconducting TMDs, *e.g.* through controlled reaction with $O_2$.[22] Recently, single-photon emission has been observed in encapsulated 2D $MoS_2$, following helium ion bombardment.[37] In this case, encapsulation removes the effects of contaminants on sulphur vacancies and other defects, while helium bombardment is believed to create Mo vacancies with localized defect states. Indeed, for single photon emission, it is not necessary for defect states to be close to the band edges, because single photons of other energies can also be considered. Our predictions also point experimentalists toward other energy ranges in which single photons may be emitted in the TMDs.

**MATERIALS AND METHODS**

Except for calculations on optical properties, all our DFT calculations are performed using the VASP code[38,39] with the PAW approach. We use the PBE exchange-correlation functional[40] with Grimme's D2[41] correction for van der Waals (vdW) interactions for all calculations except the density of states (DOS) plots which were performed with the hybrid HSE06[42] exchange-correlation functional. Spin polarization is included for the defect structures and for oxygen.

STM images were simulated in the Tersoff-Hamann approximation.[28] The Tersoff-Hamann approximation predicts STM images[14] for $S_{vac}$ in $SL-MoS_2$ in good agreement with experiment, and with more sophisticated calculations including tip effects.[43] Besides the images obtained with the PBE exchange-correlation functional (Fig. 3), we have also simulated images using the HSE06[42] functional (Fig. S15), which also agree well with experiment.

Our GW-BSE calculations are performed with Quantum-ESPRESSO[44] and BerkeleyGW[45] software. PBE optimized norm-conserving Vanderbilt (ONCV) pseudopotentials[46] were used, with a kinetic energy cutoff of 60 Ry. W semicore 5s, 5p and 5d states were included as valence electrons. Each defect is simulated in a 4×4 supercell, with atomic positions fully relaxed and with the lattice parameters fixed at 4 times of the primitive $WSe_2$ cell. We performed one-shot $G_0W_0$ calculations with a dielectric matrix cutoff of 15 Ry, slab Coulomb truncation,[47] the Hybertsen-Louie generalized plasmon pole model,[48] and the static remainder method to speed up convergence.[49] The reciprocal space was sampled using non-uniform neck subsampling (NNS) in GW and cluster sampling interpolation (CSI) in BSE.[29] Using these methods, we converged the GW quasiparticle energies in the primitive $WSe_2$ cell with 520 bands and using a 15×15×1 q-mesh with an additional 10 q-points in the Voronoi cell around q=0, which is equivalent to an effective uniform grid of more than 150×150×1 sampling. The GW gap at the K point is computed to be 2.342 eV. Keeping all parameters the same and reducing the number of bands to 92, we obtain a GW gap at K of 2.367 eV. We converged the BSE spectrum using a 24x24x1 coarse mesh interpolated onto a 72×72×1 fine mesh and the CSI scheme with sub-factor 3, which translates to an additional 10 k-points sampled of kernel elements for each cluster along the (110) direction. We checked that the spectrum is converged within 0.01 eV against a 90×90×1 fine mesh calculation. In the 4×4 supercells with defects, we used a 4×4×1 q-mesh with 10 q-point NNS, and 1500 bands to compute the GW eigenvalues, and for the BSE part, we used a 6×6×1 k-mesh interpolated onto a 18×18×1 fine mesh, together with the CSI method using sub-factor 3 for an additional 10 k-points sampled for each cluster along (110). The supercell k-mesh is converged at the same level as the primitive cell. The BSE equation is solved by direct diagonalization within the Tamm-Dancoff apporoximation (TDA), including bands with energy at least 0.5 eV beyond the VBM and CBM. The perturbative treatment of spin-orbit coupling in GW-BSE optical spectra follow the procedure in Ref. 35.

RPA calculations as well as GW-BSE with full spinor wavefunctions were performed with Yambo.[50]

STEM measurements were performed on SL-WSe$_2$, CVD grown on sapphire[27] and transferred to a TEM grid for measurement. A 60 kV accelerating voltage was used. Further details on methods used are provided in the SI.

## ASSOCIATED CONTENT

**Supporting Information:**

Further details on Methods, STM images of all intrinsic defects, STEM analysis, Exciton wavefunction analysis, GW-BSE spectra with SOC, RPA spectra for $7\times7$ supercells, and other information. This material is available free of charge *via* the Internet at http://pubs.acs.org.

## AUTHOR INFORMATION

**Corresponding authors:**

*E-mail: phyqsy@nus.edu.sg (theory); phyweets@nus.edu.sg (STM); msepsj@nus.edu.sg (STEM).

† Yu Jie Zheng and Yifeng Chen contributed equally.

**NOTES**

**Conflict of interests*:* The authors declare no competing financial interests.

**ACKNOWLEDGEMENTS**

SYQ acknowledges support from grant NRF-NRFF2013-07 from the National Research Foundation (NRF), Singapore. AW, YLH and PKG acknowledge support from ASTAR Pharos grant R-144-000-359-305. SYQ, SJP and AW acknowledge support from the Singapore NRF, Prime Minister's Office, under its medium-sized centre program. SJP is grateful to the National University of Singapore for support. SJP and PKG acknowledge MOE grant number R-144-000-389-114. Computations were performed on the CA2DM cluster and the National Supercomputing Centre (NSCC) in Singapore. YJZ acknowledges an NUS research scholarship, and discussions with Zijing Ding and Zhibo Song. We thank S. Refaely-Abramson for technical advice on using the NNS and CSI methods, and G. Eda for discussions.

# Supplementary Information for:

# Point Defects and Localized Excitons in 2D WSe$_2$

Yu Jie Zheng[1,2, †], Yifeng Chen[2, †], Yu Li Huang[1,3], Pranjal Kumar Gogoi[1], Ming-Yang Li[4], Lain-Jong Li[4], Paolo E. Trevisanutto[2], Qixing Wang[1], Stephen J. Pennycook[5,*], Andrew T. S. Wee[1,2,*], Su Ying Quek[1,2, *]

[1]Department of Physics, National University of Singapore, 2 Science Drive 3, 117551, Singapore

[2]Centre for Advanced 2D Materials, National University of Singapore, Block S14, Level 6, 6 Science Drive 2, 117546, Singapore

[3]Institute of Materials Research & Engineering (IMRE), A*STAR (Agency for Science, Technology and Research), 2 Fusionopolis Way, Innovis, 138634, Singapore

[4]Physical Sciences and Engineering, King Abdullah University of Science and Technology, Thuwal, 23955-6900, Saudi Arabia

[5]Department of Materials Science & Engineering, National University of Singapore, 9 Engineering Drive 1, Singapore 117575

†: These authors contributed equally to this work.
Corresponding Authors *: phyqsy@nus.edu.sg (theory); phyweets@nus.edu.sg (STM); msepsj@nus.edu.sg (STEM)

METHODS
1. Density Functional Theory (DFT) calculations        3
2. Chemical Vapor Deposition (CVD)       5
3. Scanning tunneling microscopy/spectroscopy (STM/S)        5
4. Scanning Transmission Electron Microscopy (STEM)        5

FIGURES

TABLES:

**METHODS**

**1. Density Functional Theory (DFT) calculations**

**Electronic Properties of Defects**

Except for the calculations for optical properties, the density functional theory (DFT) calculations in this manuscript were performed with the VASP code[1] using the PAW approach and a kinetic energy cutoff of 400 eV. We use the PBE exchange-correlation functional[2] with Grimme's D2[3] correction for van der Waals (vdW) interactions for all calculations except the density of states (DOS) plots which were performed with the hybrid HSE06[4] exchange-correlation functional. Spin polarization is included for the defect structures and for oxygen. Geometry optimization is performed with a force convergence criteria of 0.01 eV/Å for single layer (SL)-WSe$_2$, and 0.05 eV/Å for SL-WSe$_2$ supported on graphite. The lattice constant is fixed to that optimized for the pristine WSe$_2$ layer, which compares well to experimental values. The total energy of bulk WSe$_2$ is converged for the chosen energy cutoff, as well as with a Monkhorst Pack k-grid sampling of 10 × 10 × 4 in the bulk WSe$_2$ unit cell. SL-WSe$_2$ on graphite was modeled using 3 layers of graphite, with a 3 × 3 WSe$_2$ supercell on top of a 4 x 4 supercell of graphite and 13 Å of vacuum. In this supercell, the strain on WSe$_2$ was 0.61 % and the strain on graphite was -0.65 %. Defects in SL-WSe$_2$ on graphite were studied using a 2×√3$R$30° supercell of the WSe$_2$/graphite supercell described above. In these defect supercells, we used a k-mesh of 2 × 2 × 1 for geometry optimization and 4 × 4 × 1 for DOS calculations. For defects in isolated WSe$_2$, we used a 5 × 5 supercell, with a 2 × 2 k-mesh for geometry optimization and a 6 × 6 k-mesh for DOS calculations.

The formation energy of the (charge neutral) defects is defined as

$$E_f(x) = E(x) - E_{pristine} - \sum_i n_i \mu_i \qquad (1)$$

where $E(x,q)$ and $E_{pristine}$ are the total energies of the WSe$_2$/graphite supercell with and without the defect, respectively. $n_i$ denotes the number of atoms of element *i* that have been added ($n_i > 0$) or removed ($n_i < 0$), and $\mu_i$ is the chemical potential of element *i*. The chemical potentials of W and Se are linked by the stability of WSe$_2$, *i.e.*

$$\mu_{WSe_2} = \mu_W + 2\mu_{Se} \qquad (2)$$

$\mu_{WSe_2}$ is the total energy per formula unit of bulk WSe$_2$ (the conclusions are unchanged if we use instead the total energy per formula unit of monolayer WSe$_2$). $\mu_W^{max}$ and $\mu_{Se}^{max}$ are total energies per atom of bcc W metal and the molecular crystal ($R\bar{3}$ phase) of Se$_6$ molecules, respectively. The minimum of the Se chemical potentials can be derived from formula (2), *i.e.*, $\mu_{Se}^{min} = (\mu_{WSe_2} - \mu_W^{max})/2$.

For formation energies of oxygen defects in the presence of oxide precursors, we have referred the chemical potential of O to its reservoir WO$_3$ (P6mmm).

$$\mu_{WO_3} = \mu_W + 3\mu_O \qquad (3)$$

Similar formulas are used for the formation energies of the S$_{vac}$, O$_S$ and O$_{ins}$ defects in MoS$_2$. The chemical potential of Mo-rich and S-rich referred to Molybdenum (bcc) and Sulphur (alpha phase) crystals. The O chemical potential is taken to be that from MoO$_3$ (Pbnm).

For the computation of dissociation barriers and rate constants, 5×5 supercells with >15 Å vacuum were used, with a Monkhorst-Pack k-mesh of 2×2×1 and a force convergence criterion of 0.01 eV/Å. The dissociation barrier was computed using the climbing image nudged elastic band (cl-NEB) method [5]. The prefactor of the reaction was estimated within harmonic transition state theory (HTST) [6-7], and is given by $A = \frac{\prod_i^{3N} v_i^{IS}}{\prod_i^{3N-1} v_i^{TS}}$, where N is the number of free atoms, $v_i^{IS}$ and $v_i^{TS}$ are stable (real valued) normal mode frequencies at the initial states (IS) and transition states (TS) or saddle point. At the transition state, exactly one of the vibrational modes has an imaginary frequency. From the prefactor $A$, the rate constant $k$ is obtained by $k = A \exp\left[-\frac{E^{TS} - E^{IS}}{k_B T}\right]$. $E^{IS}$ and $E^{TS}$ are the energies at the initial and transition states. $k_B$ and $T$ are the Boltzmann's constant and temperature, respectively.

Besides using the PBE-D2 functional as described in the main text, we have also computed the O$_2$ dissociation barriers using the RPBE functional[8] with and without Grimme's D3 dispersion correction,[9] and the BEEF-vdW functional.[10] The RPBE functional, when applied to dissociative adsorption of O$_2$ on Cu surfaces, gave reasonable agreement with experiment,[11] while the BEEF-vdW functional performed well compared to an experimental benchmark database of molecular dissociation barriers on surfaces.[12] The results are shown in Table S1.

DFT band structures with spin orbit coupling (SOC) were also computed using VASP.

**Random Phase Approximation (RPA) calculations**

RPA calculations were performed with Yambo[13] using a damping parameter of 0.01 eV. RPA with SOC calculations were performed with spinor wavefunctions using fully relativistic ONCV pseudopotentials. For 4x4 supercells, the response function has a dimension in reciprocal space of 1000 mHa. However, for $7 \times 7$ supercells, the corresponding dimension is ~0.5 mHa due to the large computational expense.

## 2. Chemical Vapor Deposition (CVD)

Monolayer $WSe_2$ was grown directly on graphite and on sapphire substrates by chemical vapor deposition (CVD) as reported in Ref. 14. In brief, high purity $WO_3$ and Se powders are used as precursors. $WO_3$ powders were placed in a ceramic boat at the center of a furnace and Se powders were placed at the upstream side, while the graphite or sapphire substrate was positioned in the downstream side next to the $WO_3$ powders. Ar was used as carrier gas to carry the evaporated Se and $WO_3$ to the target substrates for reaction. $H_2$ was added as a reducing agent. STM and STEM confirmed that the samples used were monolayer in thickness.

## 3. Scanning tunneling microscopy/spectroscopy (STM/S)

STM/S measurements were performed in a custom-built multi-chamber system housing an Omicron LT-STM operating at ~77 K under ultrahigh vacuum conditions ($10^{-10}$ mbar). All STM images were recorded in constant current mode with tunneling current in the range 50-100 pA. Differential conductance dI/dV or STS were acquired by a lock-in amplifier with a sinusoidal modulation of 40 mV at 625 Hz. Note that the bias voltage ($V_{Tip}$) is applied on the STM tip with respect to the sample, hence negative values correspond to empty states and positive values correspond to filled states. Each STS curve was obtained by averaging hundreds of individual spectra acquired. An electrochemically etched tungsten tip was used in all measurements. Before STM investigations, the *ex-situ* grown sample was degassed at ~300 ˚C overnight to remove adsorbates (*e.g.*, $O_2$, $H_2O$, *etc.*) physisorbed during exposure in ambient conditions.

## 4. Scanning Transmission Electron Microscopy (STEM)

**Transfer to TEM grid:**

At first, the $WSe_2$/sapphire sample surface was spin-coated with poly(methyl methacrylate) (PMMA, A4, 950 K in anisole, MicroChem) at 4000 rpm for 60 second. After curing the spin-coated sample at 100° C for 10 min, the edges of the sample were scratched, so that the etching agent can easily reach the $WSe_2$-sapphire interface. The sample was then floated on NaOH (3M) etching solution to separate the PMMA coated $WSe_2$ film from the sapphire substrate. After separation, the floating film was transferred to DI water beakers a few times consecutively and later fetched using a quantifoil copper TEM grid. The TEM grid with PMMA-coated $WSe_2$ film was thereafter heated at 100° C for 5 min to get better adhesion and to remove water. The PMMA coating was washed off finally, using acetone and IPA. This transfer process was carried out inside a class 1000 cleanroom in Singapore Synchrotron Light Source (SSLS) situated at the National University of Singapore (NUS) campus.

**Scanning Transmission Electron Microscopy:**

Aberration-corrected scanning transmission electron microscopy (STEM) images were taken at 60 kV accelerating voltage using the JEOL ARM200 F installed inside SSLS, NUS. The microscope, with 80 pm resolution at 200 kV and demonstrated information transfer of 95 pm at 40 kV, is capable of high-resolution imaging to reveal the atomic structure and defects. The lower accelerating voltage of 60 kV used here for $WSe_2$ is much below the knock-on damage threshold of Se and W atoms in $WSe_2$, and hence it ensures negligible beam-damage. High-angle annular dark-field (HAADF) image intensities depend on the atomic number of the corresponding atoms in the sample. All the images reported here were taken with the HAADF detector with the collection angle range of 68-280 mrad. The convergence angle of the probe beam was about 30 mrad.

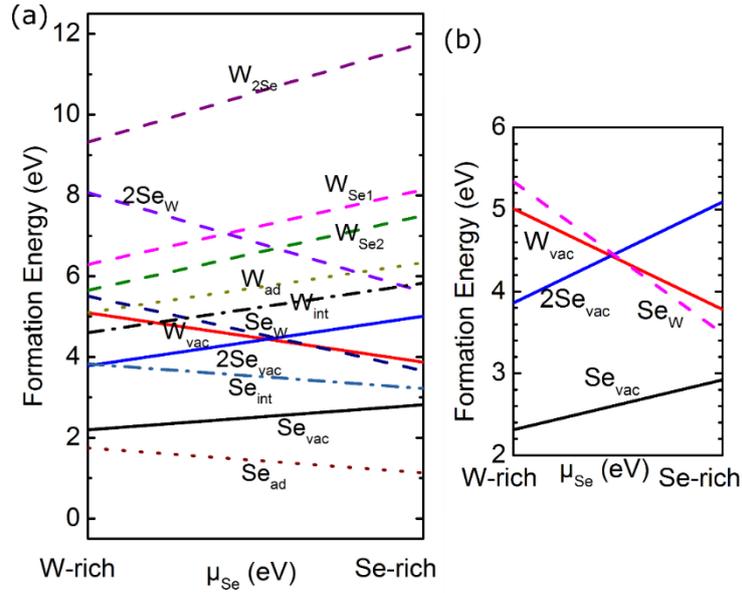

FIG S1. Formation energies of intrinsic defects (a) with and (b) without graphite substrate. Solid lines denote vacancies, dashed lines substitutional (antisite) defects, dotted lines adatoms, and dotted-dashed lines intercalated atoms.

| Defect | W-rich conditions | | Se-rich conditions | |
|---|---|---|---|---|
| | Isolated | Supported | Isolated | Supported |
| $Se_{vac}$ | 2.31 | 2.20 | 2.92 | 2.82 |
| $W_{vac}$ | 5.01 | 5.09 | 3.78 | 3.87 |
| $Se_W$ | 5.34 | 5.51 | 3.49 | 3.66 |
| $2Se_{vac}$ | 3.86 | 3.78 | 5.09 | 5.01 |
| $O_{Se}$ | 0.52 | 0.33 | 0.73 | 0.53 |
| $O_{ins}$ | 2.38 | 2.36 | 1.97 | 1.95 |
| $O_{ad}$ | 2.91 | 2.81 | 2.50 | 2.40 |

Table S1. Formation energies in eV of selected intrinsic defects, and O-related defects in isolated WSe$_2$ ML and in WSe$_2$ ML supported on graphite.

| Defect | Mo-rich conditions | S-rich conditions |
|---|---|---|
| $S_{vac}$ | 1.55 | 2.81 |
| $O_S$ | -0.09 | 0.33 |
| $O_{ins}$ | 3.72 | 2.88 |

Table S2. Formation energies in eV of selected defects in isolated $MoS_2$ ML.

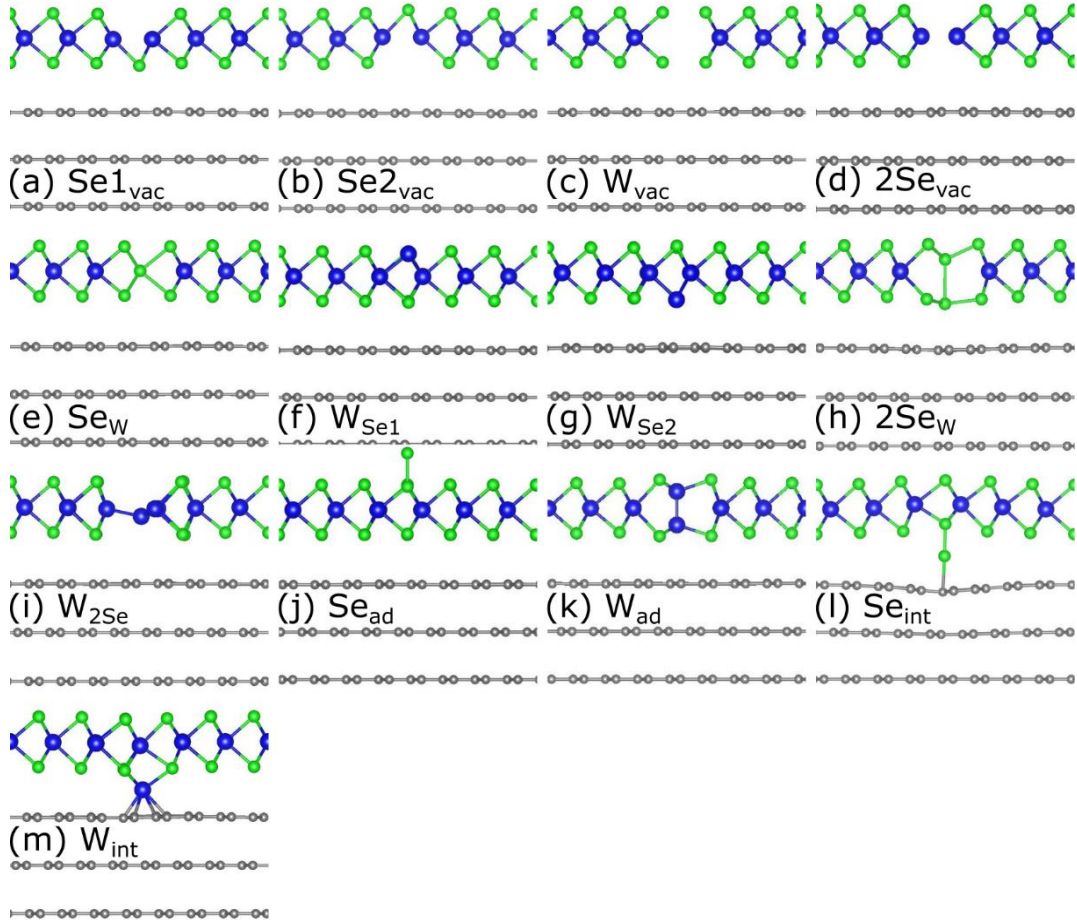

FIG. S2. Side view of the intrinsic point defects in $WSe_2$ ML on graphite. Blue: W, Green: Se, Gray: C. We note that $W_{Se2}$ has a slightly smaller formation energy than $W_{Se1}$. Analysis of the detailed atomic structure indicates that the W-W bonds are ~0.1Å longer in $W_{Se2}$ than in $W_{Se1}$, suggesting that interaction with graphite stabilizes the substitutional W atom.

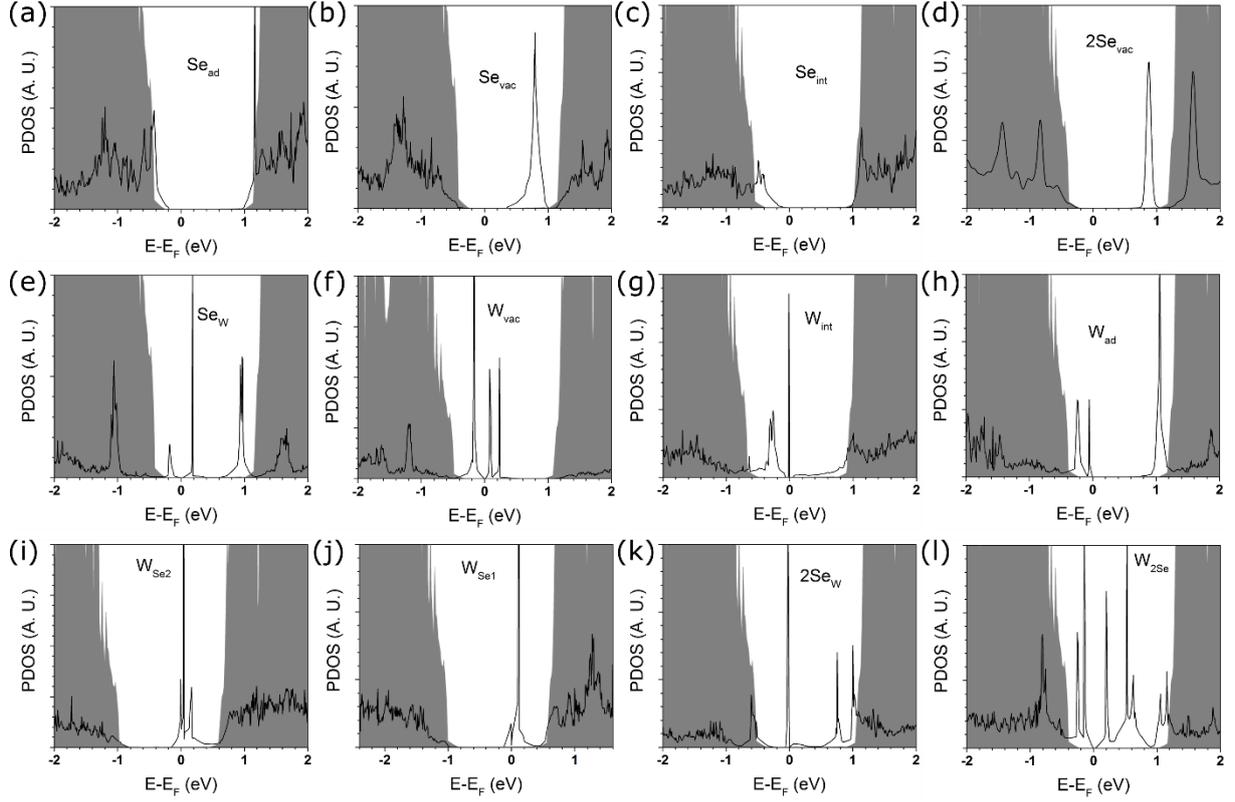

FIG. S3. PDOS (DFT PBE) on the intrinsic defect sites in WSe$_2$ ML on graphite. Gray shading: PDOS of perfect WSe$_2$ on graphite. The PDOS with and without defects are aligned by the 1s core level of W furthest from the defect site. The Fermi level refers to the Fermi level of WSe$_2$ ML with defects on graphite.

Table S3. Energy barrier for dissociation of $O_2$ on $Se_{vac}$ in $WSe_2$, computed for different exchange-correlation functionals.

|  | PBE-D2 | PBE-D2 (with zero point correction) | RPBE | RPBE-D3 | BEEF-vdW[1] |
|---|---|---|---|---|---|
| Barrier (eV) | 0.52 | 0.48 | 0.54 | 0.57 | 0.51 |

[1.] For BEEF-vdW, the number in the table is obtained from structures where the lattice constant is fixed to that of PBE-D2, which is closer to experiment. Using the lattice constant optimized with BEEF-vdW (overestimated by ~3.5%), we obtain a much smaller energy barrier of 0.l3 eV.

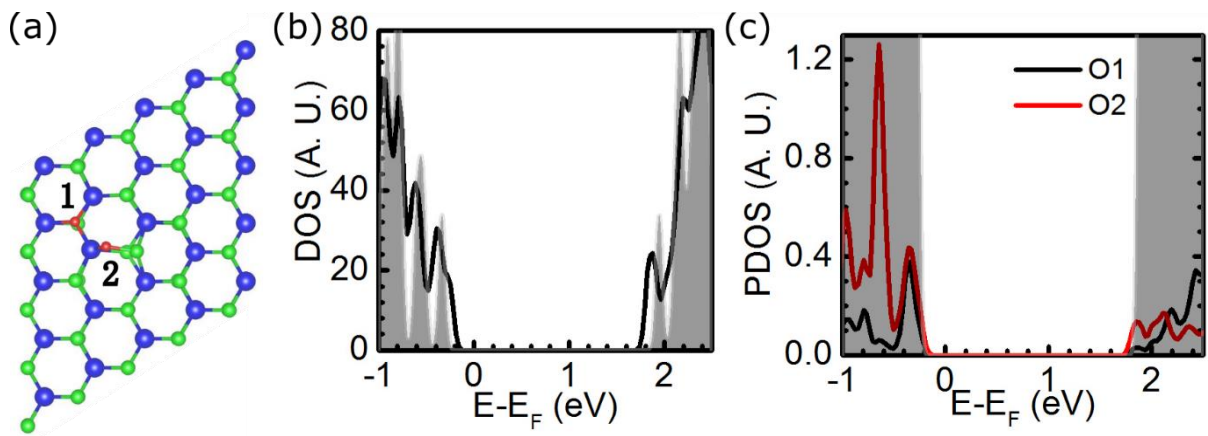

FIG. S4. Atomic structure and DOS for the final state in Figure 2a of the main text. (a) Top view of atomic structure (Blue: W, Green: Se, Red: O), (b) DOS of final state, (c) PDOS on the O atoms which are marked in (a). Gray shading represents the DOS of perfect $WSe_2$, aligned using the 1s levels of W atoms furthest from the defect.

Table S4. Binding energies ($E_b$) in eV of the O atoms in the final state. $E_b = E_{2O} - E_{1O} - \mu_O$, where $E_{2O}$ is the total energy of the final state in Figure 2a, $E_{1O}$ is the total energy after the specified O atom is removed, and $\mu_O$ is half of the total energy of O$_2$, and for the number in brackets, $\mu_O$ is the energy of atomic O. Spin polarization is included. We see that the O1 atom is much more strongly bound that O2.

|  | O1 | O2 |
| --- | --- | --- |
| $E_b$ | -4.2 (-6.8) | 0.3 (-2.3) |

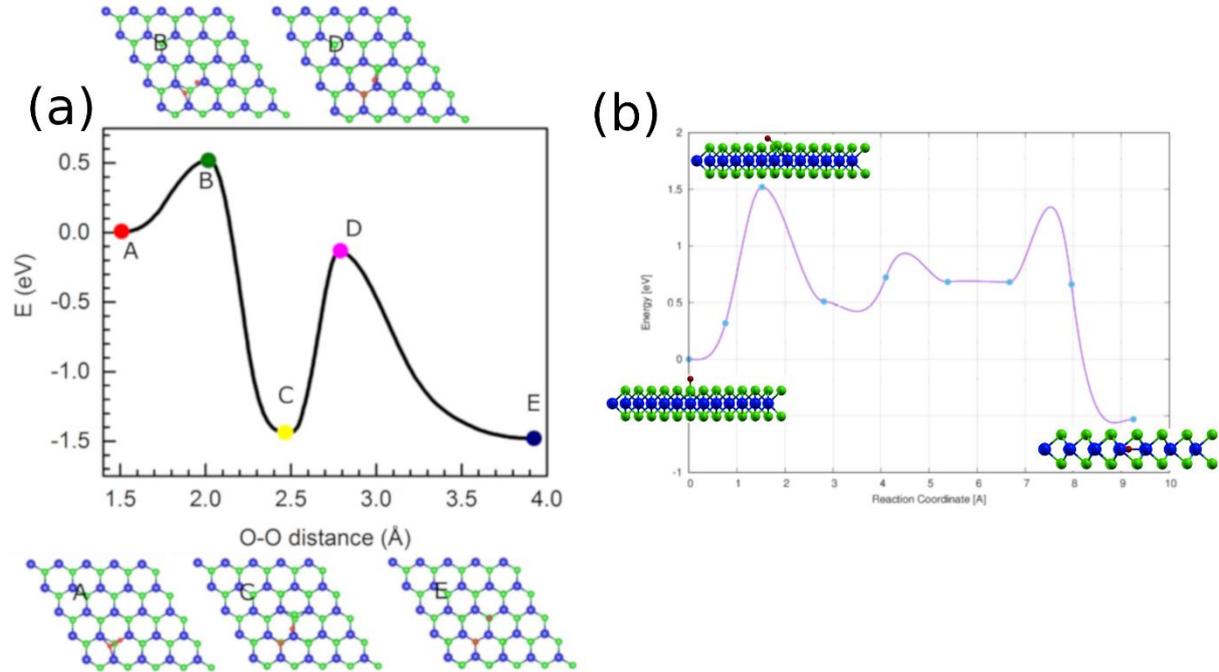

FIG. S5. (a) O$_2$ dissociative adsorption at the Se vacancy site and diffusion of oxygen atom to (a) O$_{ad}$; (b) Energy barrier for O atom to diffuse from O$_{ad}$ to O$_{ins}$. The energy barrier from O$_{ad}$ to O$_{ins}$ is 1.5 eV, smaller than the binding energy of O atoms (relative to atomic O) at the different binding sites.

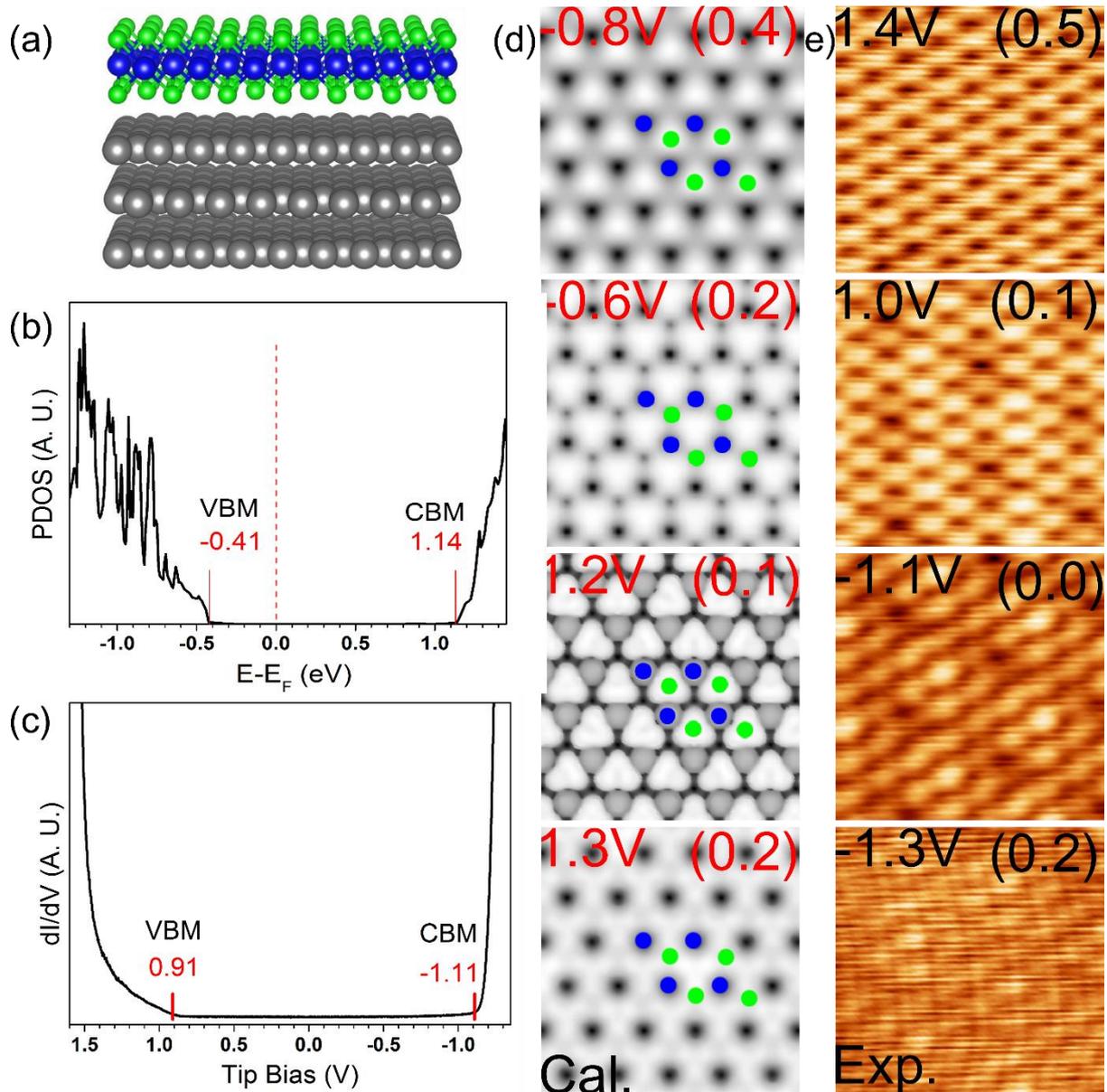

FIG. S6. STM/S results for perfect SL-WSe$_2$ on graphite. (a) Atomic structure. (Blue: W; Green: Se; Gray: C) (b) PDOS (PBE) of WSe$_2$ on graphite reveals a 1.55 eV band gap. (c) STS spectrum reveals a 2.02 eV bandgap for SL-WSe$_2$ ($V_{Tip}$=1.3 V, 68.5 pA) on graphite with slight p-type doping characteristic, similar to previous reports.[15] (d) Simulated STM images at different bias voltage (values in parenthesis referenced to the band edges) with atoms overlain in the images (Blue spots: W, Green spots: Se). Energy ranges chosen for the simulation approximate those in the experiment. (e) Bias dependent experimental STM images of SL-WSe$_2$ on graphite (values in parenthesis referenced to band edges).

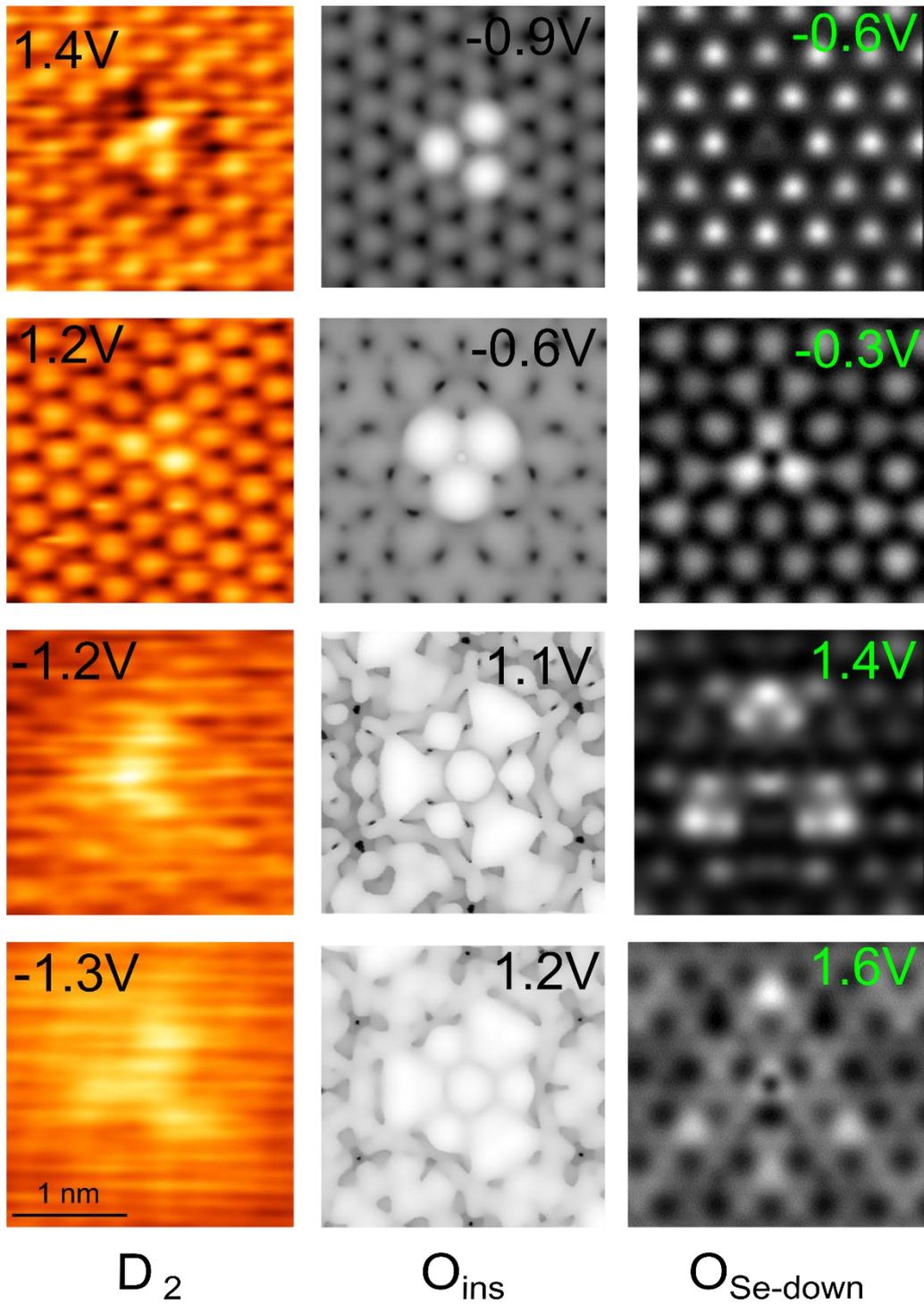

FIG. S7. STM images for $D_2$, compared with simulated STM images for $O_{ins}$ and $O_{Se-down}$ (O substituting the bottom Se atom)

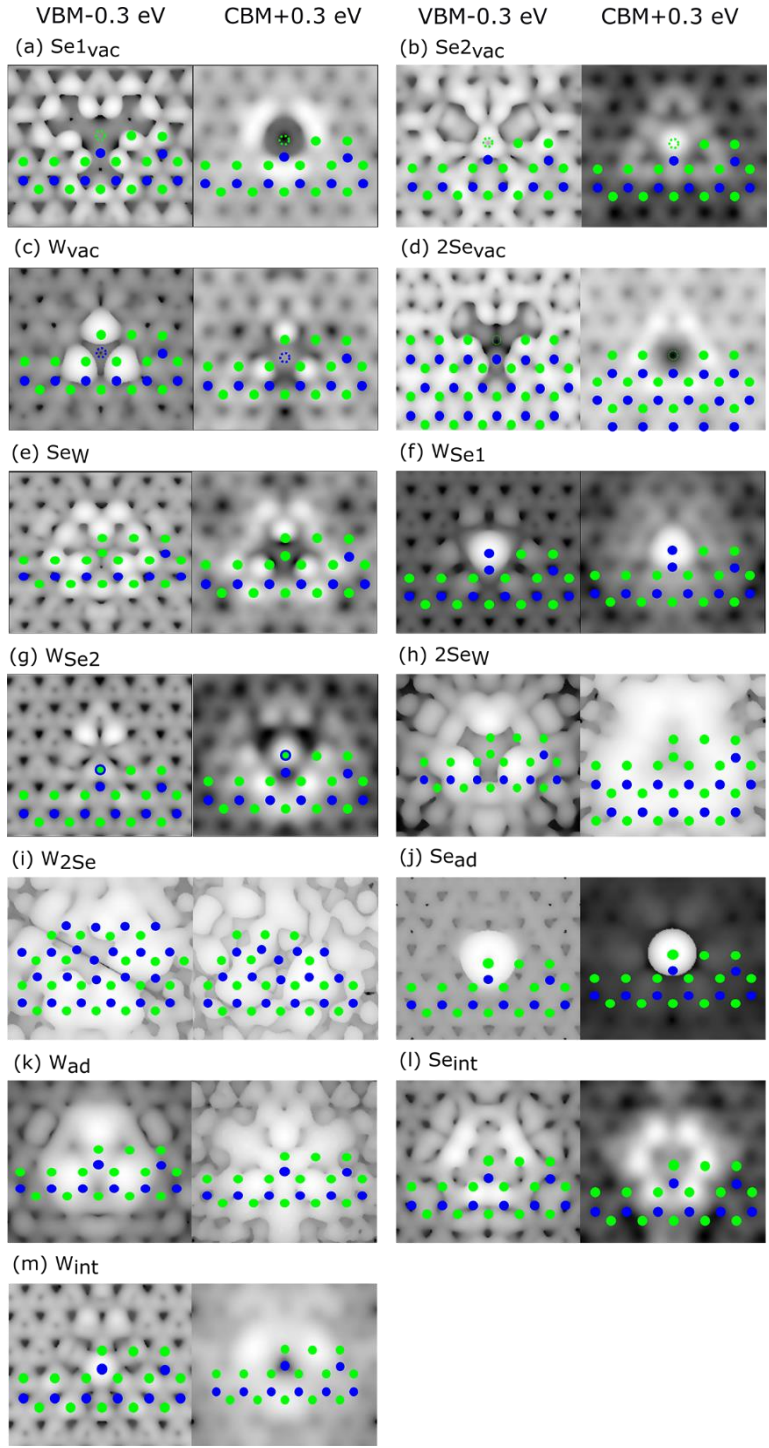

FIG. S8. Simulated STM images of the intrinsic defects in SL-WSe$_2$ on graphite. The energy ranges are chosen to probe occupied and unoccupied states that are 0.3 eV away from the band edges. Atomic positions are overlain on the images. Blue: W, Green: Se. Se1$_{vac}$ and Se2$_{vac}$ refer to single Se vacancies with the missing Se being in the top and bottom layer, respectively.

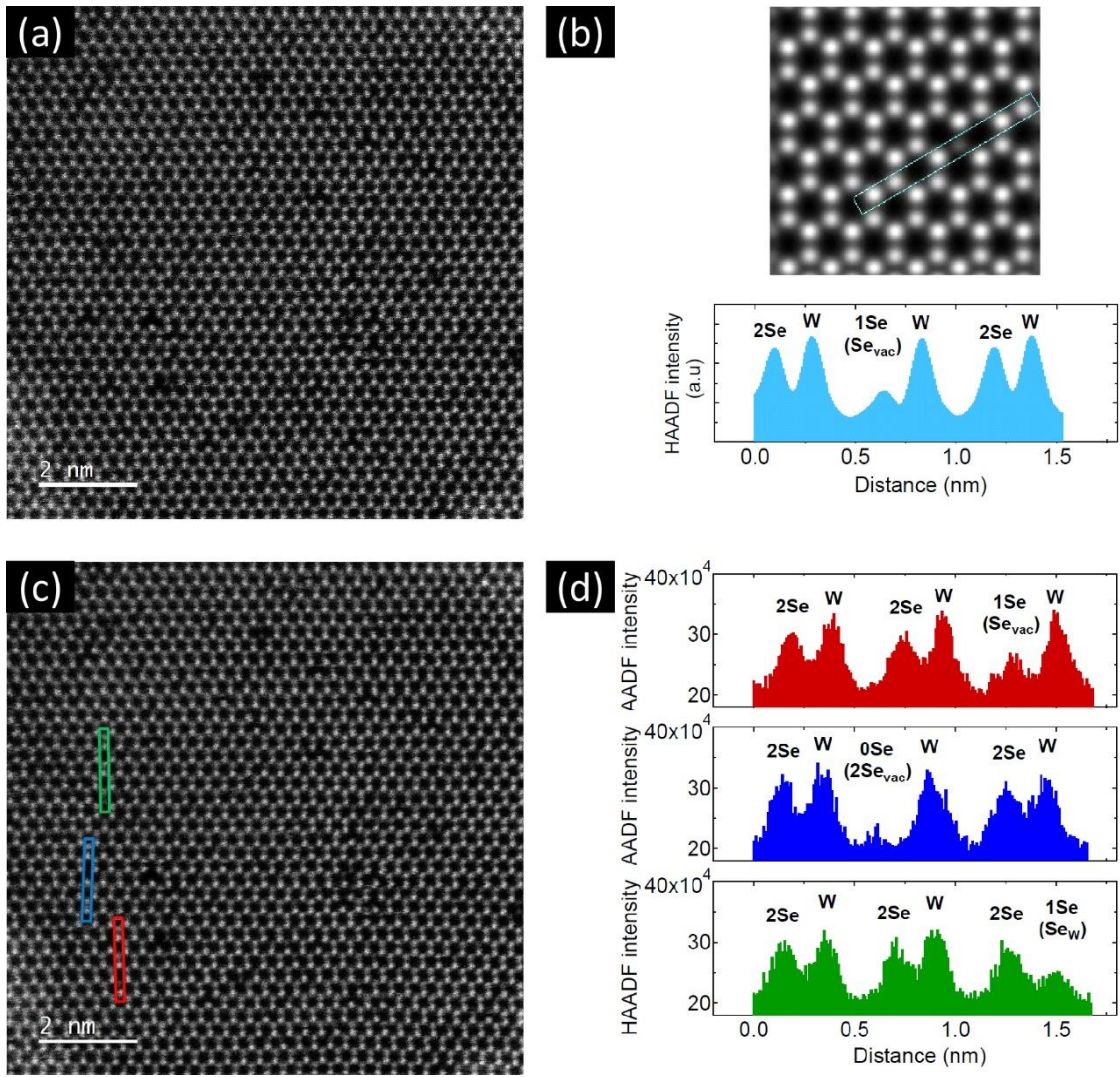

FIG. S9. HAADF STEM image at 60 kV accelerating voltage together with intensity analysis. (a) Raw experimental HAADF STEM image presented in the main text Fig. 4 (g) and (h) (no overlaid atoms now) is shown. (b) Simulated HAADF STEM image and intensity profile, obtained using QSTEM simulation package with input parameters same as in experiments, are shown for the representative case of a region with one $Se_{vac}$. The brighter spots are due to W atoms and the lesser intensity ones are due to two Se atoms at the Se site. One $Se_{vac}$ is in the simulated image as can be seen in the line profile. The $Se_{vac}$ can be easily distinguished both in the simulated image as well in the line profile due to its reduced intensity. (c) The same experimental HAADF STEM image as in (a) is shown now with line profiles indicated. The red line profile captures a $Se_{vac}$, the blue line profile captures a $2Se_{vac}$, and the green line profile captures a $Se_W$. (d) The corresponding intensity line profiles are plotted for the regions indicated in (c). These experimental line profile intensities (taking the background also into consideration) are used to distinguish the nature of defects by comparing them with corresponding simulated image intensities quantitatively.

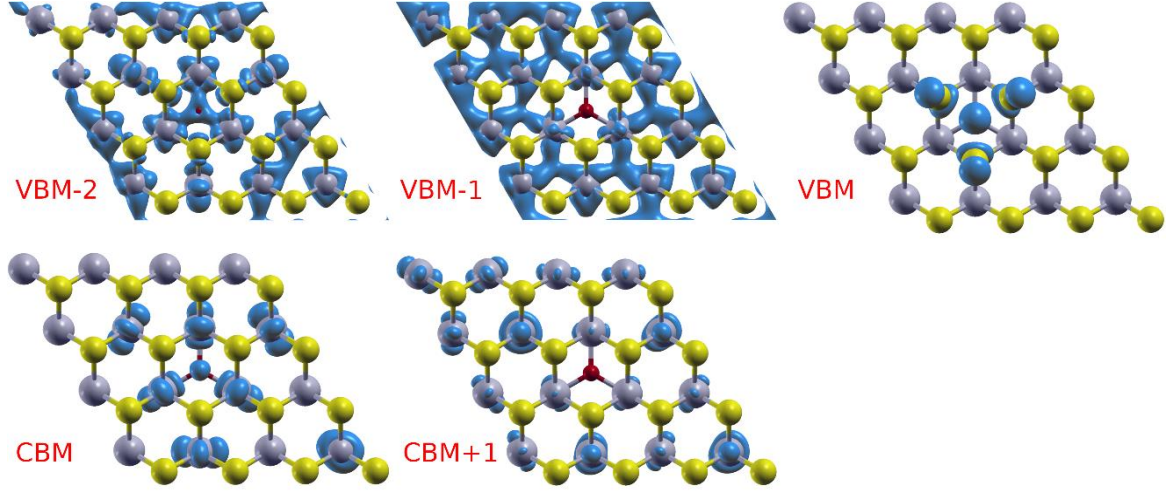

FIG. S10. Charge density plots for DFT eigenfunctions at the K point, in the 4 x 4 supercell with $O_{ins}$. The isovalue is set to 10% of the maximum of the respective charge distribution, except for VBM, which is set to 1% for better visualization.

| LX1: 1.783 eV | CBM | CBM+1 | CBM+2 | LX2: 1.860 eV | CBM | CBM+1 | CBM+2 |
|---|---|---|---|---|---|---|---|
| VBM | 0.00006 | 0.00000 | 0.00000 | VBM | 0.00002 | 0.00001 | 0.00000 |
| VBM-1 | <span style="color:red">0.88401</span> | 0.04781 | 0.00009 | VBM-1 | <span style="color:red">0.41199</span> | 0.11755 | 0.00366 |
| VBM-2 | 0.02064 | 0.00593 | 0.00002 | VBM-2 | <span style="color:red">0.40391</span> | 0.00617 | 0.00269 |
| LX3: 1.880 eV | CBM | CBM+1 | CBM+2 | A: 1.978 eV | CBM | CBM+1 | CBM+2 |
| VBM | 0.00065 | 0.00002 | 0.00000 | VBM | 0.00036 | 0.00015 | 0.00002 |
| VBM-1 | <span style="color:red">0.66412</span> | 0.08961 | 0.00271 | VBM-1 | 0.23457 | <span style="color:red">0.42535</span> | 0.01096 |
| VBM-2 | 0.17760 | 0.00832 | 0.00434 | VBM-2 | 0.19933 | 0.02762 | 0.01315 |

Table S5: Analysis of exciton wavefunction for first few bright exciton states in the 4x4 supercell with $O_{ins}$. The numbers shown are $\sum_{\vec{k}} A^S_{vc\vec{k}}$ for each combination of valence state $v$ and conduction state $c$, where the exciton wavefunction is given as $\Psi^S(\vec{r}_e, \vec{r}_h) = \sum_{vc\vec{k}} A^S_{vc\vec{k}} \psi_{c\vec{k}}(\vec{r}_e) \psi_{v\vec{k}}(\vec{r}_h)$. Major contributing entries are highlighted in red. It is important to note that the GW self-energy corrections result in a reordering of the valence states at this K point. In Table S5, all states are labeled according to the energy ordering in the mean-field DFT (PBE) calculation. At the K point, the DFT states VBM, (VBM-1) and (VBM-2) correspond, respectively, to the GW (VBM-2), GW VBM and GW (VBM-1).

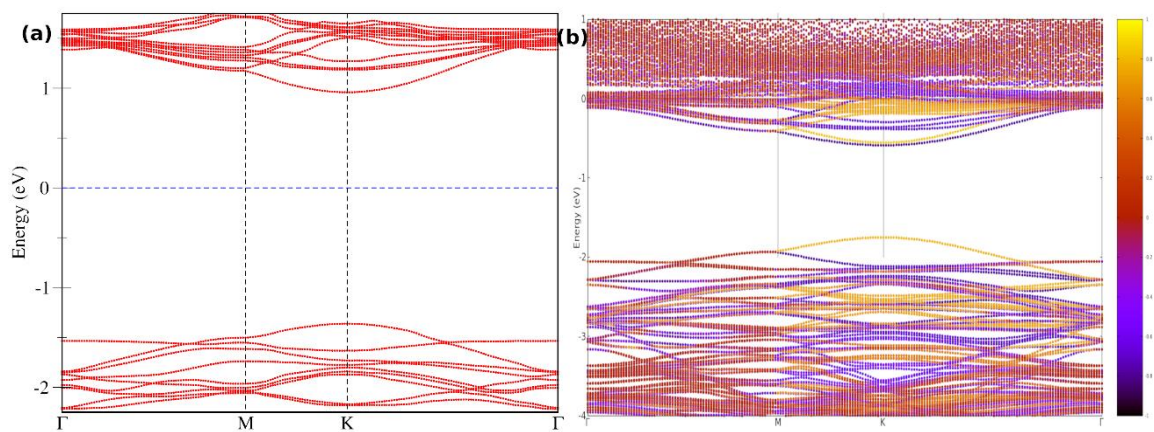

FIG S11. (a) GW and (b) DFT-SOC band structure with $m_z$ projection for 4x4 supercell with $O_{Se}$ defect.

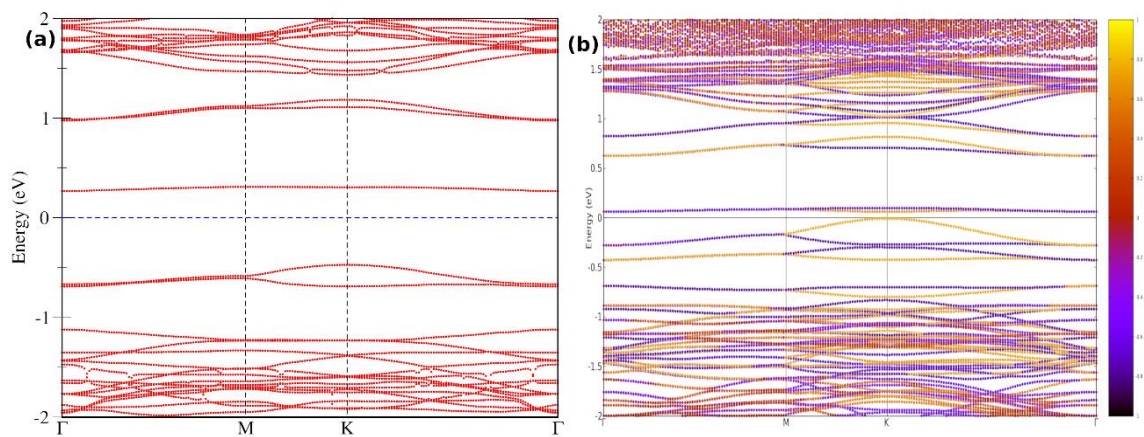

FIG S12. (a) GW and (b) DFT-SOC band structure with $m_z$ projection for 4x4 supercell with $Se_W$ defect.

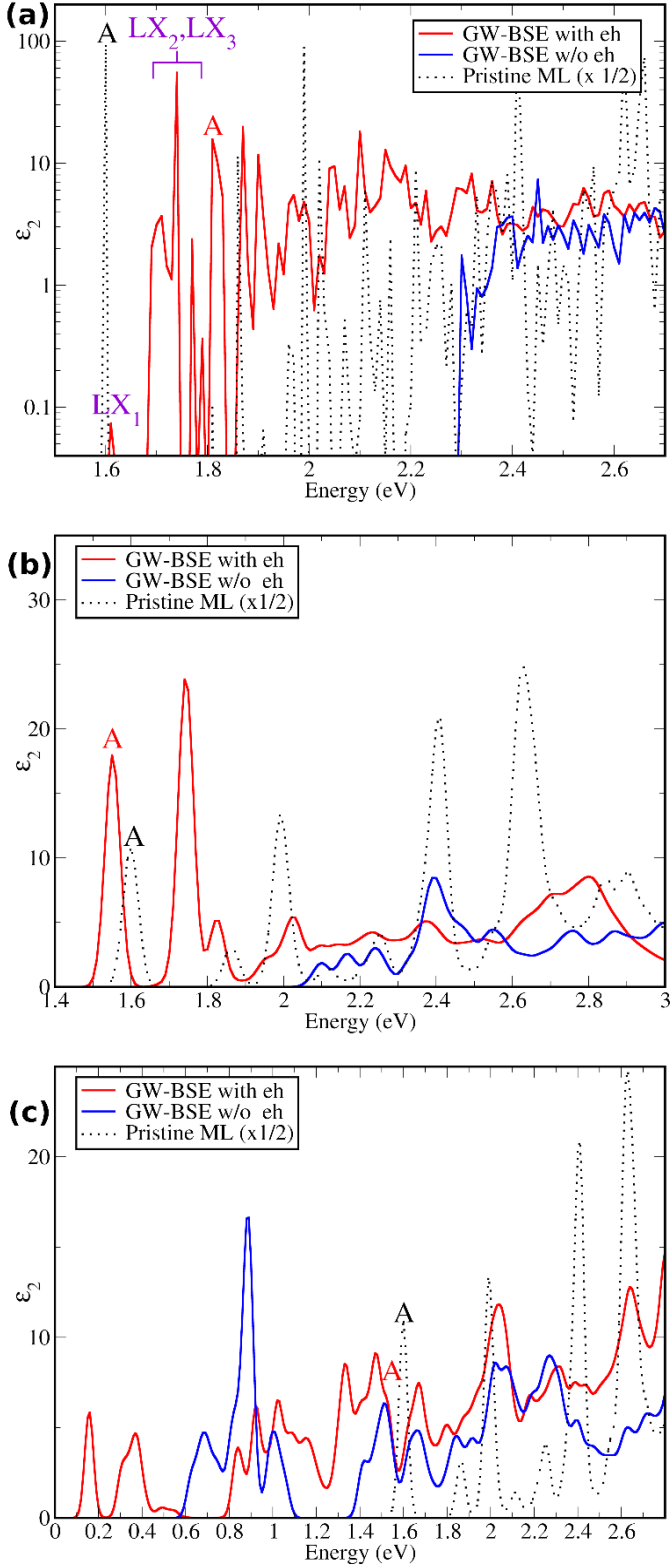

FIG. S13. GW-BSE absorption spectra with spin orbit coupling (SOC) applied perturbatively, for 4x4 WSe$_2$ supercell with (a) O$_{ins}$, (b) O$_{Se}$ and (c) Se$_W$ defect. SOC is also applied perturbatively for the independent particle absorption spectra (blue) and for the pristine WSe$_2$ monolayer (black dashed). We note that one should focus only on peaks with energies not larger than the free exciton A peaks in each system.

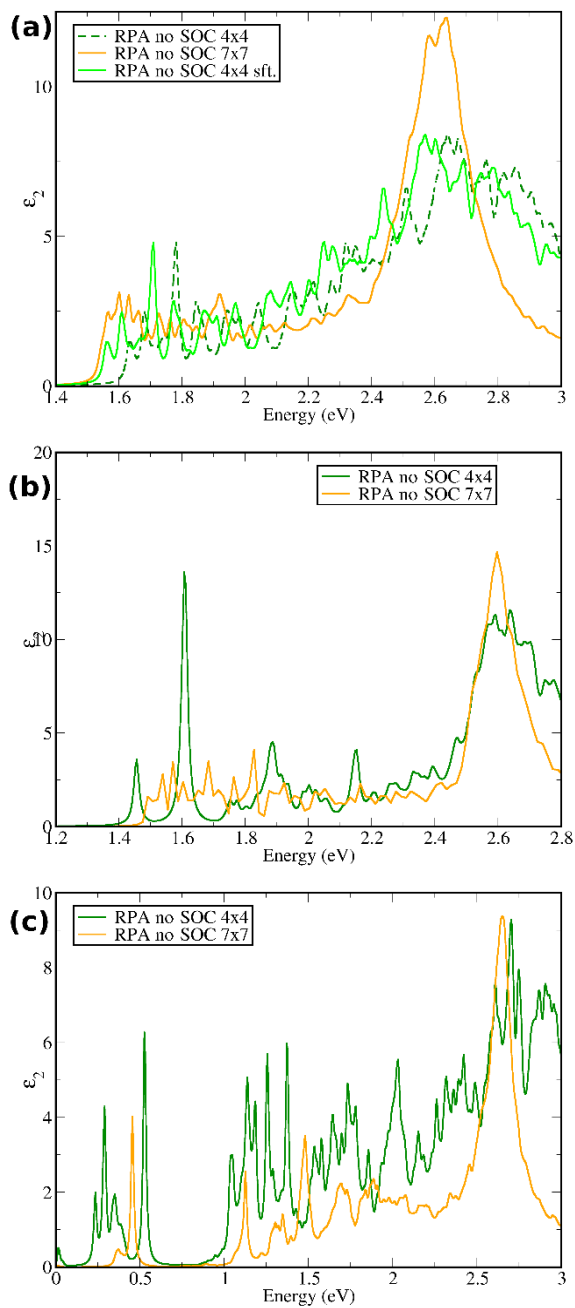

Figure S14. RPA spectra without SOC for 7x7 supercell of (a) O$_{ins}$, (b) O$_{Se}$, and (c) Se$_W$ anti-site defect, in comparison with the corresponding 4x4 supercell result. The 0.072 eV red shifted 4×4 spectrum in (a) is to account for the DFT gap size difference between 4×4 and 7×7 supercell systems. The fact that fewer peaks are present in the 7×7 supercells may be related to the limited number of valence and conduction bands in these calculations.

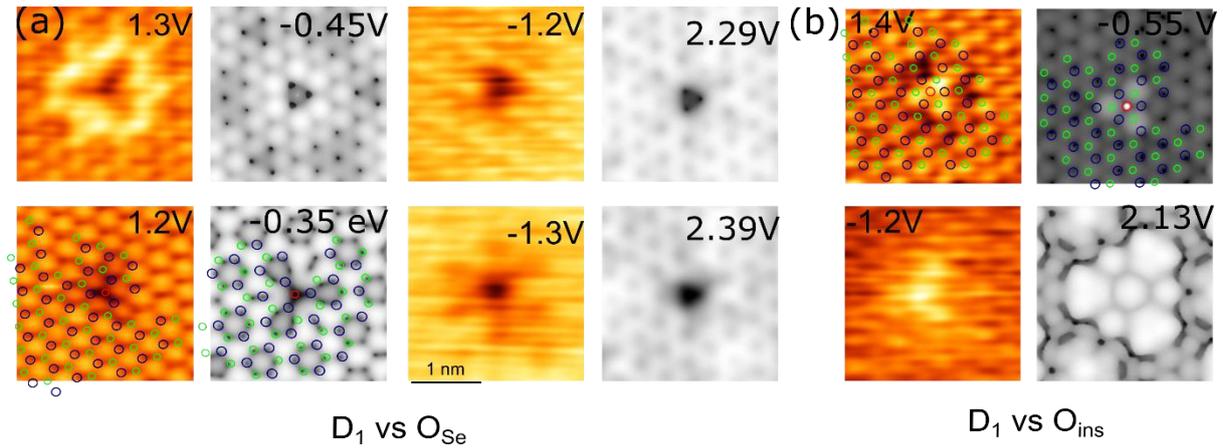

FIG. S15. STM images simulated with HSE06 functional. Left: Experimental STM (Omicron LT-STM operating at ~77 K and $10^{-10}$ mbar) images for (a) $D_1$ and (b) $D_2$. Right: Simulated STM images of (a) $O_{Se}$ and (b) $O_{ins}$ using the HSE06 exchange-correlation functional. The atoms are overlain in the images for D1 and D2 defects. (Red: O; Blue: W; Green: Se).

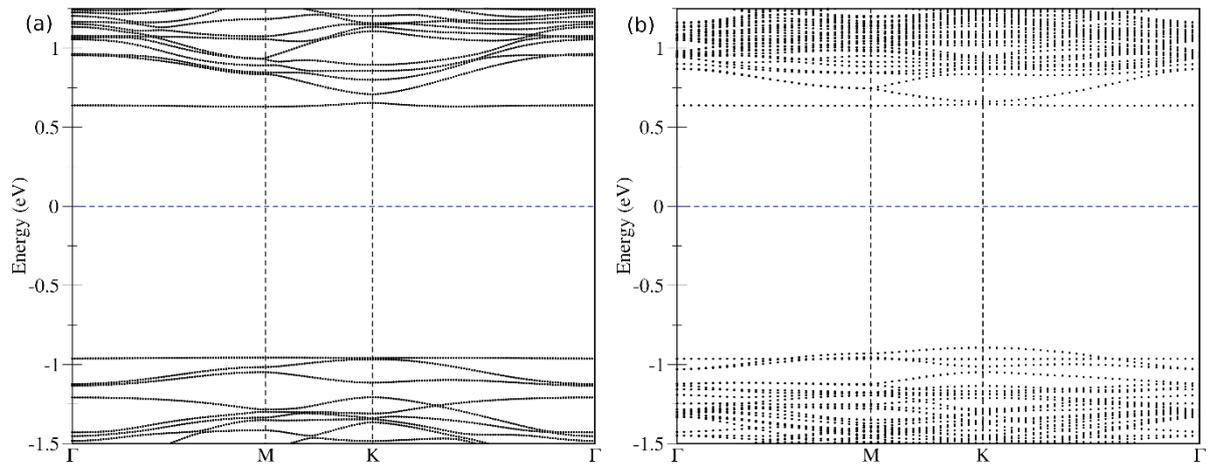

FIG. S16. DFT-PBE band structure of $O_{ins}$ defective system in (a) 4x4 supercell and (b) 7x7 supercell.

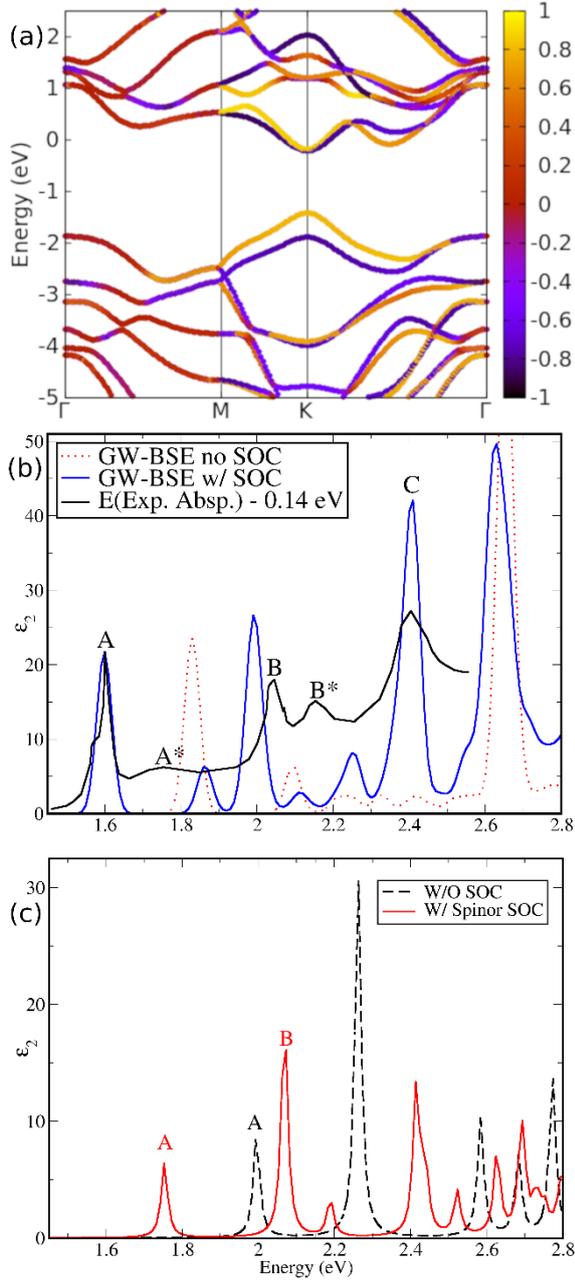

FIG. S17. SOC effects on the electronic and optical spectra of pristine WSe$_2$ (a) PBE-SOC bandstructure of primitive cell pristine monolayer WSe$_2$, with m$_z$ value encoded in the color depth scale. (b) GW-BSE optical absorption spectra of pristine WSe$_2$ without and with perturbative SOC, in comparison with experimental spectrum (from BerkeleyGW). (c) GW-BSE optical spectra of pristine WSe$_2$ without and with full spinor SOC from Yambo code. Note that here the two spectra are blue shifted rigidly by the same 0.49 eV amount to match with the experimental A peak value at 1.75 eV, due to the underlying under-converged GW quasiparticle spectrum for these calculations.